\documentclass[journal]{IEEEtran}

\usepackage{geometry}
\usepackage{helvet}
\usepackage{xspace}
\usepackage{url}
\usepackage{enumitem}
\usepackage[hidelinks]{hyperref}
\usepackage[utf8]{inputenc}
\usepackage{graphicx} 
\usepackage{xcolor}
\usepackage{indentfirst} 
\usepackage{tabularx} 
\usepackage{wrapfig}

\usepackage[export]{adjustbox}

\usepackage{subfig}
\usepackage[compact]{titlesec}
\titlespacing{\section}{0pt}{1ex}{0.5ex}
\titlespacing{\subsection}{0pt}{0.5ex}{0ex}
\titlespacing{\subsubsection}{0pt}{0.5ex}{0ex}

\newif\ifdraft
\drafttrue 

\usepackage{comment}

\newcommand{\pengcheng}[1]{\ifdraft{\textcolor{magenta}{\textbf{PW:} #1}}\fi}
\newcommand{\somali}[1]{\ifdraft{\textcolor{red}{SC: #1}}\fi}

\newcommand{\edgardo}[1]{\ifdraft{\textcolor{blue}{\textbf{EB:} #1}}\fi}

\newcommand{\newtext}[1]{\ifdraft{\textcolor{red}{#1}}\else{#1}\fi}
\newcommand{\rev}[1]{\ifdraft{\textcolor{blue}{#1}}\else{#1}\fi}

\newcommand{\ie}{{\em i.e.,}\xspace}

\newcommand{\eg}{{\em e.g.,}\xspace}

\newcommand{\customsec}[1]{\noindent{\textbf{{#1}:}}}

\newcommand{\name}{{\sc Orpheus}\xspace}

\usepackage[font={bf,footnotesize},labelsep=colon,justification=justified]{caption}

\title{\name: Living Labs for End-to-End Data Infrastructures for Digital Agriculture}
\author{Pengcheng Wang, Edgardo Barsallo Yi, Tomas Ratkus, Somali Chaterji}
\date{September 2021}

\begin{document}

\maketitle

\begin{abstract}
IoT networks are being used to collect, analyze, and utilize sensor data. There are still some key requirements to leverage IoT networks in digital agriculture, e.g., design and deployment of energy saving and ruggedized sensor nodes (SN), reliable and long-range wireless  network  connectivity,  end-to-end  data  collection pipelines  for  batch  and  streaming  data.  Thus,  we  introduce  our  living  lab ORPHEUS and  its  design  and  implementation  trajectory  to showcase  our  orchestrated  testbed  of  IoT  sensors,  data connectivity,  database  orchestration,  and  visualization dashboard.  We deploy light-weight energy saving SNs in the field to collect data,  using  LoRa  (Long  Range  wireless)  to  transmit  data from  the  SNs  to  the  Gateway  node,  upload  all  the  data to  the  database  server,  and  finally  visualize  the  data.  For  future  exploration,  we  also  built  a testbed of embedded devices using   four   different   variants   of   NVIDIA   Jetson   development  modules  (Nano,  TX2,  Xavier  NX,  AGX  Xavier) to  benchmark  the  potential  upgrade  choices  for  SNs  in ORPHEUS.  Based  on  our  deployment  in  multiple  farms in  a  3-county  region  around  Purdue  University,  and  on the  Purdue  University  campus,  we  present analyses from our living lab deployment and additional components of the next-generation IoT farm. 
\end{abstract}


\section{Introduction}
\label{sec:intro}
The promise of Digital Agriculture is to feed the growing population of the world with  a smaller footprint---less irrigation, agrochemicals, and space. There has long been talk of using computing technologies to make that promise a reality. Yet, this has not come to pass for several reasons, three of which are relevant to our discussion here. \textit{First}, the sensors to capture fine-grained spatial and temporal data are either not available, are expensive, or are error-prone. \textit{Second}, the ML platforms are not flexible enough to cater to the varied needs of the different farms. \textit{Third}, the computing equipment is not available, such as to run fast analytics at the edge, and the wireless communication is unreliable, sparse, or not available at the bandwidth required.

In the digital agriculture model, equipments such as field cameras, temperature sensors, humidity monitoring sensors, and aerial drones are deployed for data acquisition. These generate streaming, often real-time data, which can be harnessed to guide production decisions, support preventive maintenance of equipment, develop intelligent logistics, and diversify risk management methods, thereby optimizing the efficiency of resource allocation. There is still a large room for improvement, considering the relative lag of agricultural digitization vis-\`a-vis scarce resources for feeding the world. 

\begin{figure*}[htbp]
    \centering
    \includegraphics[width=0.85\textwidth]{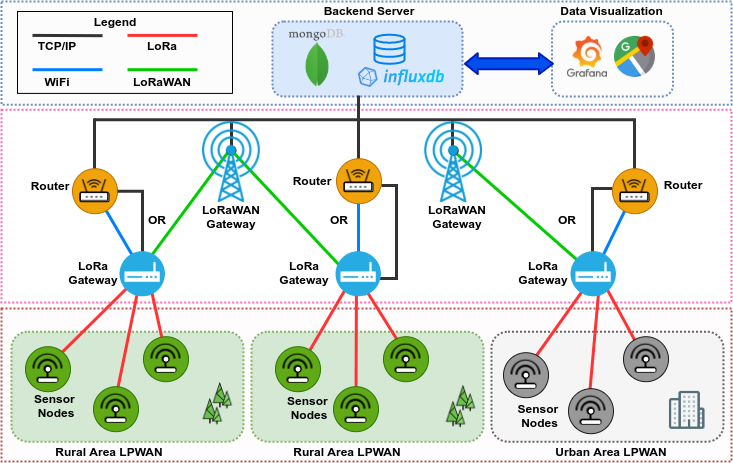}
    \caption{
    {Three-layer IoT network topology of our living testbed: SNs, Gateway, and the backend servers. The SNs and LoRa Gateways form an LPWAN using LoRa wireless connection at the edge. 
    The LoRa Gateways upload the data collected from the SNs to the backend over TCP/IP, or forward it to a nearby LoRaWAN Gateway, depending on the availability of Internet access.}
    Finally, data visualization is built using a Grafana dashboard based on the data at the backend server.}
    \label{fig:topology}
    \vspace{-1em}
\end{figure*}

In \name, we describe an end-to-end production pipeline of sensing-actuation, transport, and analytics to enable timely actions in digital agriculture toward sustainable agricultural practices. It brings together deploying low-cost energy-saving sensors in a ``living lab'' setting, wireless networking and information theory, applied data analytics, and domain expertise in digital agriculture. We demonstrate \name in operational farm settings --- at Purdue's experimental farms (Agronomy Center for Research and Education (ACRE) and Throckmorton Purdue Agricultural Center (TPAC)) and commercial production farms in counties in the vicinity of Purdue, such as Benton and Warren counties. Our vision is logically organized into three categories --- {\bf Sense} (sensor deployment in the living lab environment, Fig.~\ref{fig:nitrate},~\ref{fig:soil-2}), {\bf Transport} (wireless networking, Fig.~\ref{fig:topology}), and {\bf Analyze} (anomaly detection notifications, database aggregation, and dashboard for visualization, Fig.~\ref{fig:grafana},~\ref{fig:website},~\ref{fig:pdr})\footnote{We make a subset of the data collected from our living lab testbed available for broad research use at~\cite{purduewhin-sensor-list}. We exclude all data that is privately owned and released by the data owner.}. 


\noindent In \name, we make the following contributions:
\begin{enumerate}[leftmargin=1em]
    \item We have deployed a living testbed with dozens of low-power sensors, deployed in a 3-county region, 
    in northern and central Indiana. We show an end-to-end IoT network building scheme in real farms. See our network topology in Fig.~\ref{fig:topology}. Our network consists of: the edge part that uses SNs (sensor networks, \ie network of embedded devices with sensors) to collect data, and the Gateway nodes, which gather data from all the local SNs. Then, gateway(s) will send all the data to a central server where we have a frontend to visualize the data. We use long-range wireless communication (specifically LoRa) to send data from the SNs to the gateway, which can be miles apart, forming an LPWAN (low-power wide area network or simply LPN, \ie low-power network). Our SNs are power-saving and field-environment resistant, allowing for collection of data over months using standard batteries. Our experience is that these SNs last for 3-5 months on 4 AAA batteries, which we plan on reducing further using our data reduction algorithm, as described in our system---Ambrosia~\cite{suryavansh2021ambrosia}.
    We call this real IoT network a \textit{living testbed} for use in downstream applications, such as approximate computer vision applications to monitor crop health, or livestock monitoring.
    \item We also built an embedded testbed in our lab that consists of System-on-Chips (SoCs) from NVIDIA, used for benchmarking various lightweight IoT and object detection protocols~\cite{chaterji2020cloud, emdl2021} for use in surveillance in our experimental farms, \eg in livestock monitoring~\cite{suryavansh2021ambrosia}.  
    We leverage heterogeneous edge devices, as follows, in our testbed: \textit{NVIDIA Jetson Nano}, which has 4GB memory, and a 128-core Maxwell GPU; \textit{NVIDIA Jetson TX2}, which has 8GB memory, and a 256-core Pascal GPU; \textit{NVIDIA Jetson Xavier NX}, which has a 8GB memory, and a 384-core Volta GPU with 48 Tensor cores; and NVIDIA Jetson AGX Xavier, which has 32GB memory and 512-core Volta GPU. These Jetson-class devices~\cite{nano,tx2,xaviernx,agxxavier} feature compute-constrained, ML-capable GPUs, and range in price from \$100 to \$700.
    We use this testbed to perform computer vision tasks ranging from object classification to object detection to apply to digital agriculture scenarios. Further, drones can be used as data ferries for opportunistically offloading data from the SNs to the gateway nodes in the case of sparse or intermittent connectivity or for offloading large volumes of data. In addition, drones can be equipped with additional embedded devices as payloads to do custom-object detection using our specialized embedded computer vision software~\cite{emdl2021}, such as ApproxNet for object classification~\cite{xu2019approxnet} and ApproxDet for object detection~\cite{xu2020approxdet}.
    \item We bring out insights about the three aspects of IoT---\textit{Sense, Transport, and Analyze}---realized through actual (farm) deployments. These insights can serve to enable further efforts both in research and in development of production IoT networks. 
    \end{enumerate}
\section{Implementation of the Living Testbed of IoT Nodes}
\label{sec:living-testbed}
{Here, we describe the components and technologies to set up the living testbed. First, we describe each of the components in Fig.~\ref{fig:topology}, followed by how these components are integrated into our three-layer architecture.}

\subsection{{Components of sensor nodes (SNs)}}
\label{subsec:sensor-node}
\begin{figure}[htb]
    \centering
    \includegraphics[width=0.9\columnwidth]{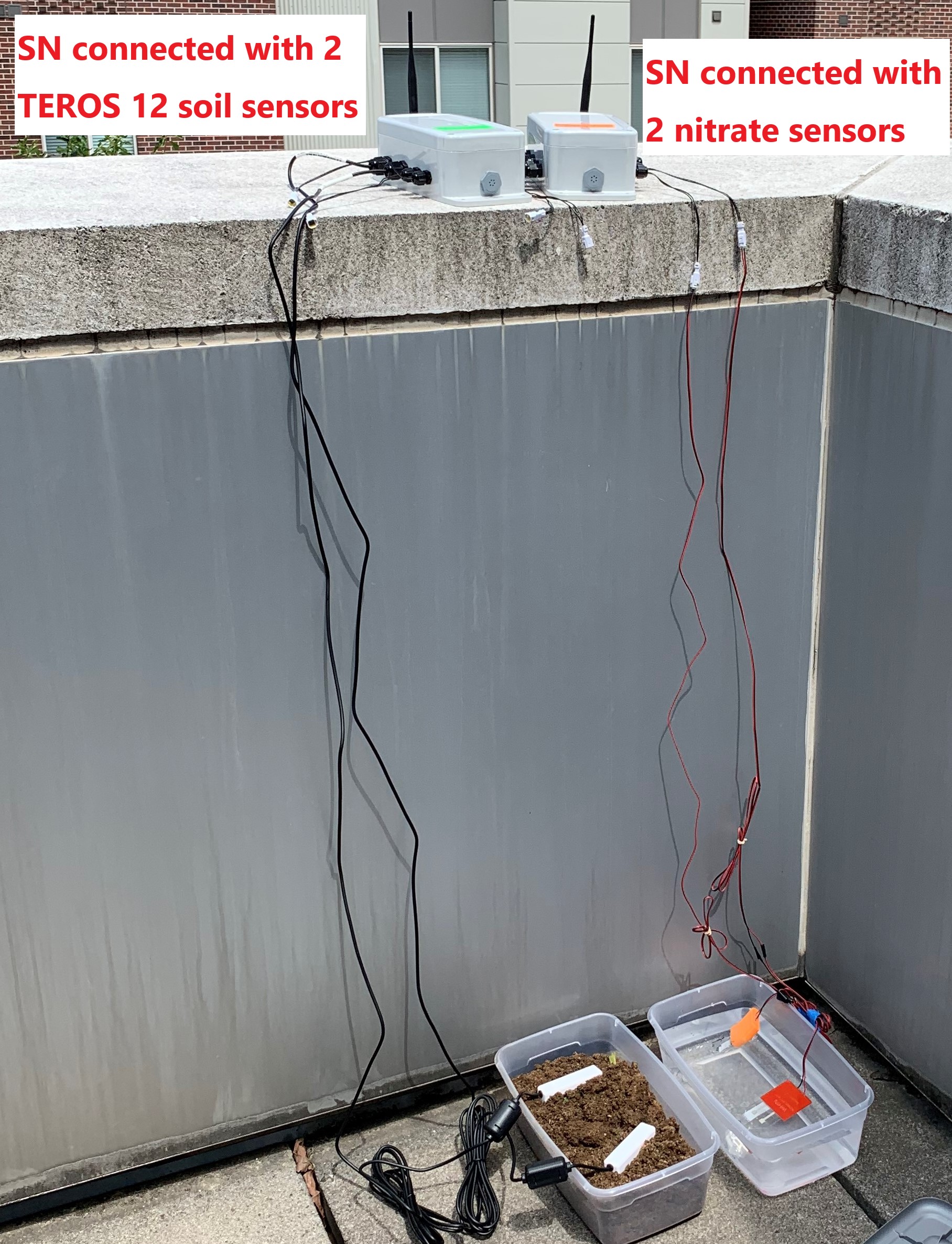}
    \caption{Packaged SNs connected with soil, and nitrate sensors. Location: Purdue University main campus.}
    \label{fig:node}
\end{figure}

\begin{figure}[htb]
    \centering
    \includegraphics[width=\columnwidth]{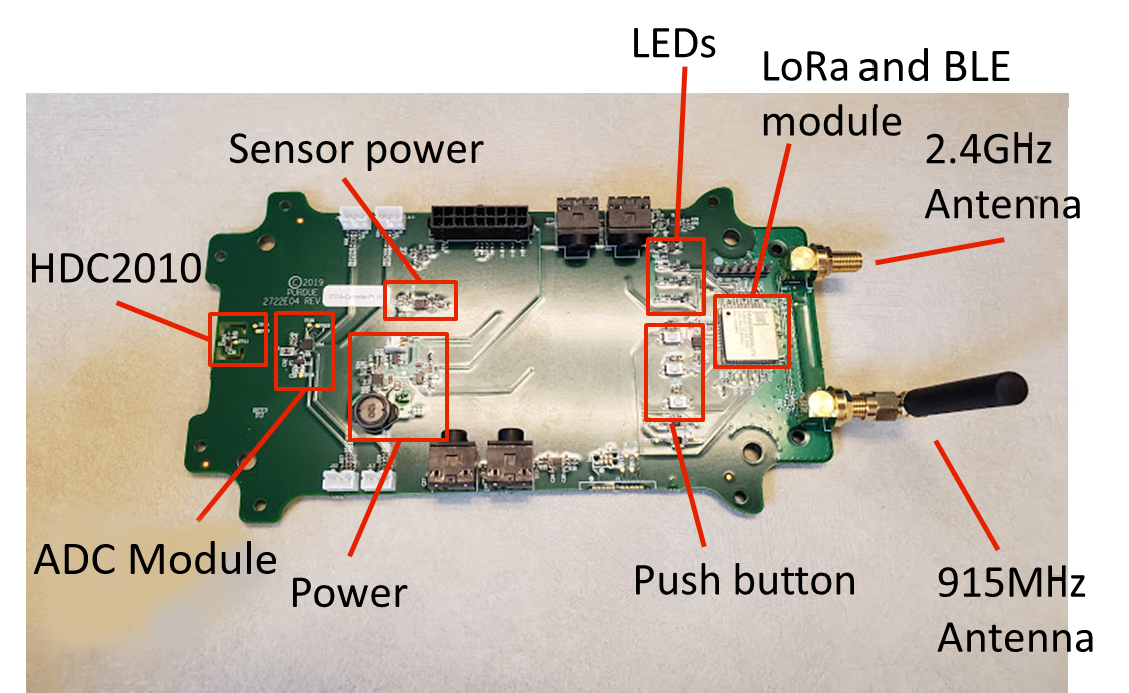}
    \caption{Embedded board (sensor node) with LoRa and BLE modules, on-board temperature \& humidity sensors (HDC2010), 2 antenna ports---the 915MHz port for LoRa and the 2.4GHz port for BLE---8 sensor ports to connect with external sensors, the ADC module converts the analog signal from nitrate sensors to a digital signal, and the LEDs are for functional checking.}
    \label{fig:pcb}
\end{figure}

\begin{figure}[htb]
    \centering
    \includegraphics[width=\columnwidth]{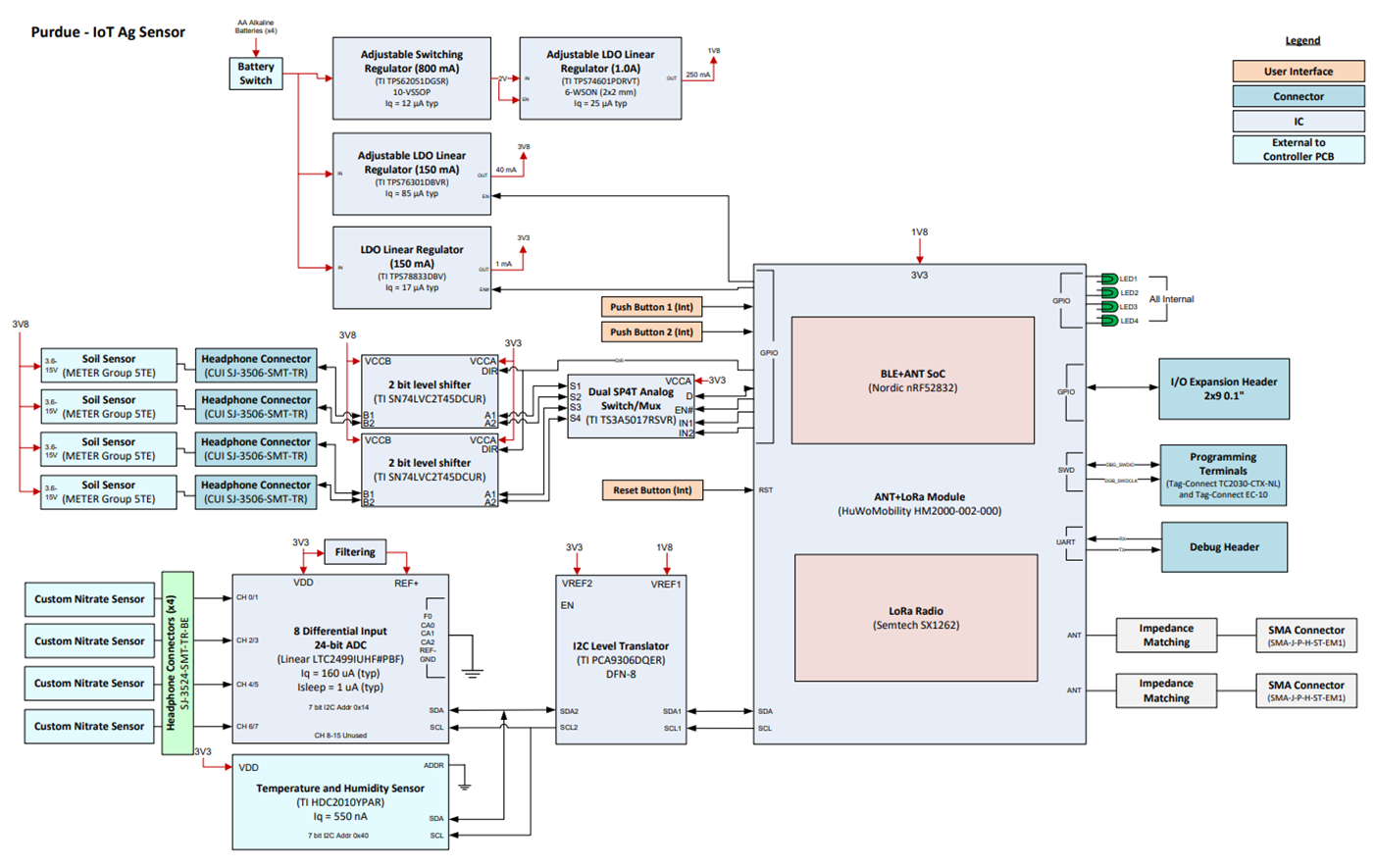}
    \caption{Block diagram of the embedded board for SN.}
    \label{fig:pcb_design}
\end{figure}

\noindent{\customsec{Embedded Boards for our Digital Agriculture Application}}
The following features are considered while designing the SN (shown in Fig.~\ref{fig:node}) for our living lab: 1) Durability; 2) Energy efficiency; 3) Ruggedized packaging.
Fig.~\ref{fig:pcb} shows the printed circuit board (PCB) used for our SN. The most important component is the HM200 module from HuWoMobility~\cite{HuWoMobility}. It includes a Nordic nRF52832 SoC~\cite{nRF52832} (with a 512 KB Flash and 64 KB RAM) and a Semtech SX1262 chip~\cite{SX1262}, allowing both Bluetooth Low Energy (BLE) and LoRa wireless communication protocols, respectively. Fig.~\ref{fig:pcb_design} shows the block diagram of the custom PCB used in our testbed. 
SX1262 has the following features: a) designed for a long battery life---only 4.2 mA of active current consumption; b) transmits up to +15dBm, with a highly efficient integrated power amplifier; c) supports LoRa modulation and LoRaWAN protocols, which makes the model configurable to satisfy different application demands; d) continuous frequency coverage from 150MHz to 960MHz allows the support of all major sub-GHz ISM bands around the world. Besides the microcontroller module, our SN includes an on-board temperature and humidity sensor HDC2010~\cite{hdc2010} from Texas Instruments, one antenna port for LoRa, and one port for BLE.
In addition, 8 sensor ports can be connected with 4 nitrate sensors and 4 soil sensors.

\customsec{On-board Temperature \& Humidity Sensor}
The HDC2010 is an integrated humidity and temperature sensor that offers high precision measurements with low power consumption in an ultra-compact Wafer-Level Chip Scale Package. The HDC2010 includes a heating element to dissipate condensation and moisture, improving its reliability. It is also resilient against dirt, dust, and other environmental contaminants. Moreover, the HDC2010 digital features provide programmable interrupt thresholds that support alerts and device wakeups, without needing a microcontroller to continuously track the system. The HDC2010 supports operation from -40°C to +125°C, with a temperature accuracy of 0.2°C, and 0\% to 100\% relative humidity with an accuracy of 2\%, which could satisfy a vast number of environmental monitoring IoT applications.

\begin{figure}[htbp]
    \centering
    \begin{minipage}[t]{0.47\columnwidth}
    \centering
    \includegraphics[width=\textwidth]{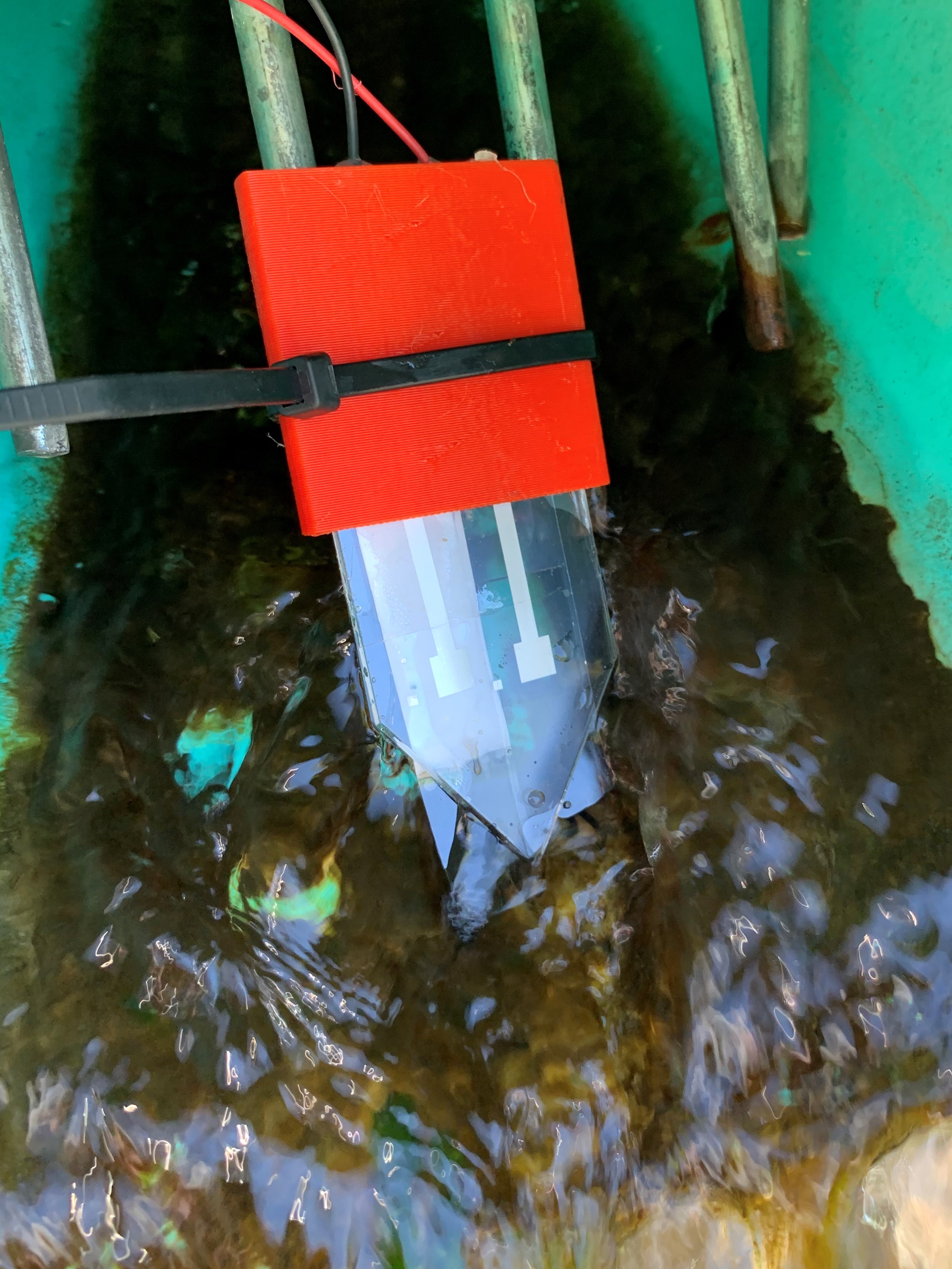}
    \caption{Nitrate sensor of a sensor node in a farm drain tile. A tile drainage system consisting of a network of underground pipes allowing subsurface water to move out from between soil particles into the tile line.}
    \label{fig:nitrate}
    \end{minipage}
    \hspace{0.02\linewidth}
    \begin{minipage}[t]{0.47\columnwidth}
    \centering
    \includegraphics[width=\textwidth]{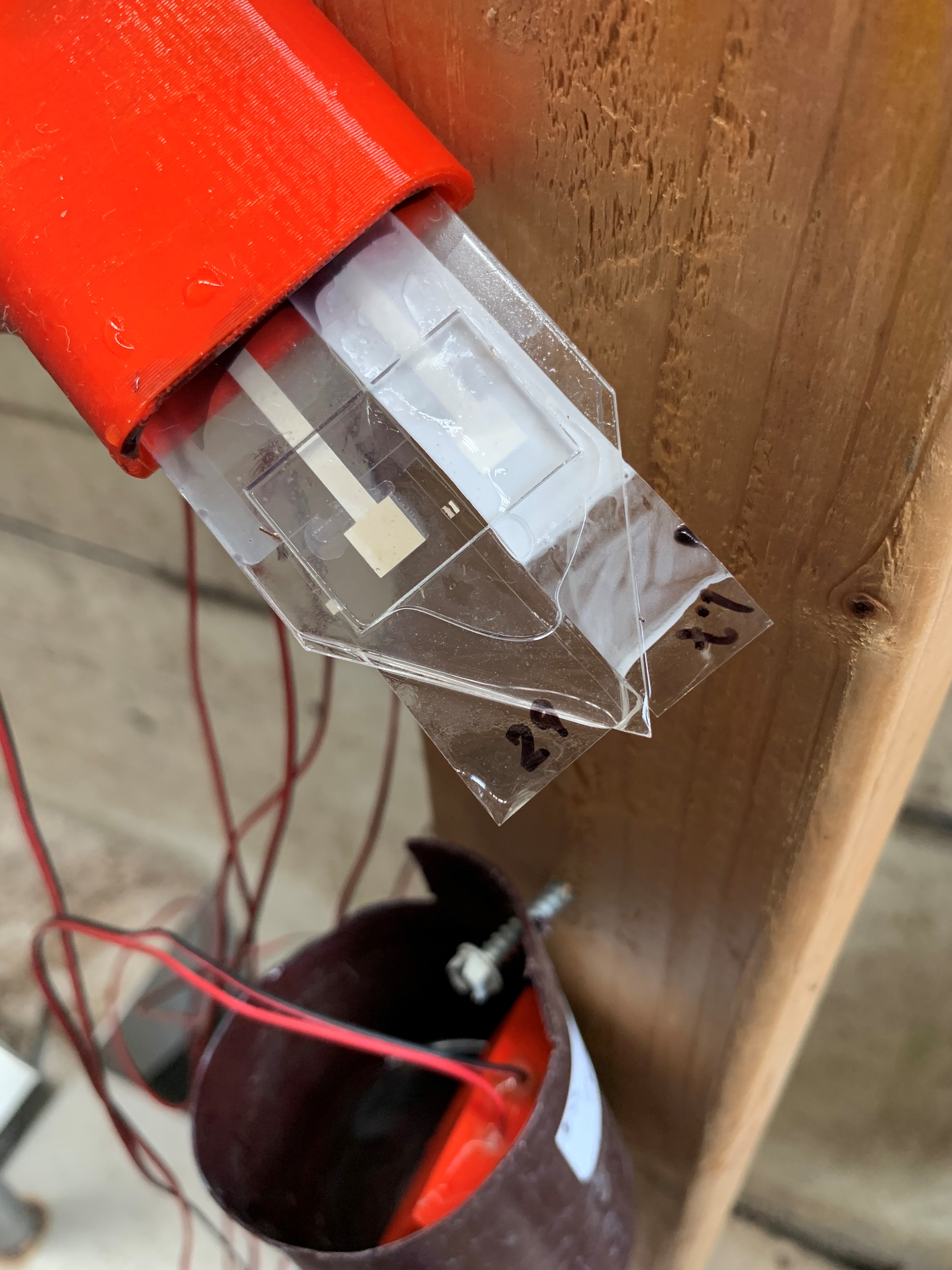}
    \caption{Nitrate sensor of a sensor node in a water quality field station~\cite{wqfs} at Purdue's experimental farm---Agronomy Center for Research and Education (ACRE). The water is collected using a drainage system.}
    \label{fig:nitrate-2}
    \end{minipage}
\end{figure}

\noindent{\textbf{Nitrate Sensor:}}
Our node can be connected with, at most, 4 (in-house developed) nitrate sensors (see Fig.~\ref{fig:nitrate} and Fig.~\ref{fig:nitrate-2}), and 4 soil moisture sensors, respectively. 
The nitrate sensors are flexible, screen-printed, detecting nitrate levels in the water. The sensors are comprised of reference and sensing electrodes. The electrodes are fabricated by printing silver ink onto a polyethylene terephthalate substrate. The sensing electrode is coated with an ion-selective membrane, which is selective to nitrate ions, and the rest of the sensor is passivated with a silicone solution. The reference electrodes are passivated at the sensing area. The difference between the nitrate-sensitive electrode and a reference electrode is used to determine the nitrate concentration~\cite{jiang2019wireless}.



\begin{figure}[htbp]
    \centering
    \begin{minipage}[t]{0.47\columnwidth}
    \centering
    \includegraphics[width=\textwidth]{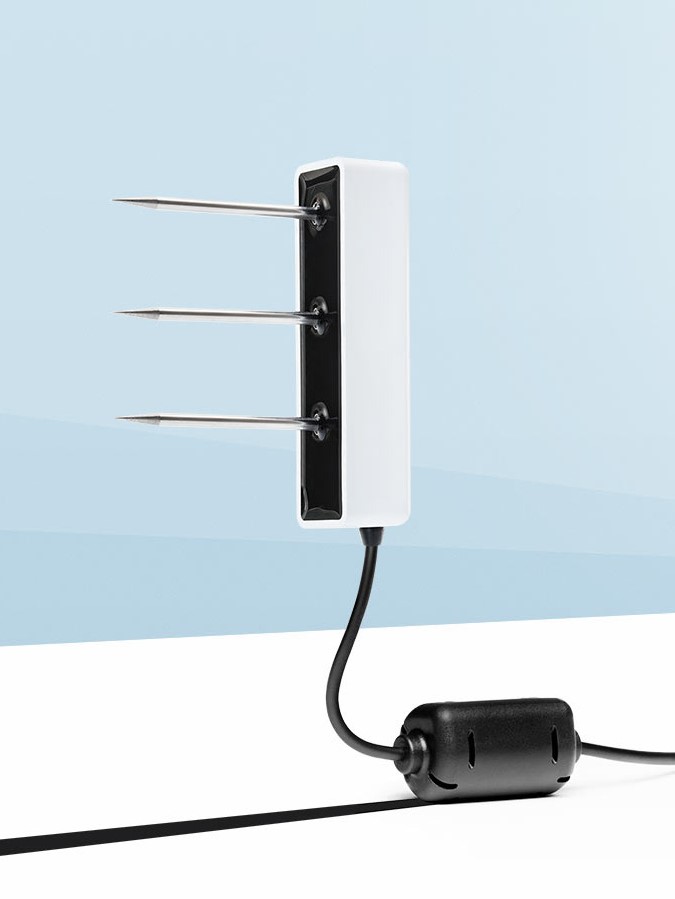}
    \caption{TEROS 12 Soil Moisture Sensor from METER~\cite{meter}.}
    \label{fig:soil}
    \end{minipage}
    \hspace{0.02\linewidth}
    \begin{minipage}[t]{0.47\columnwidth}
    \centering
    \includegraphics[width=\textwidth]{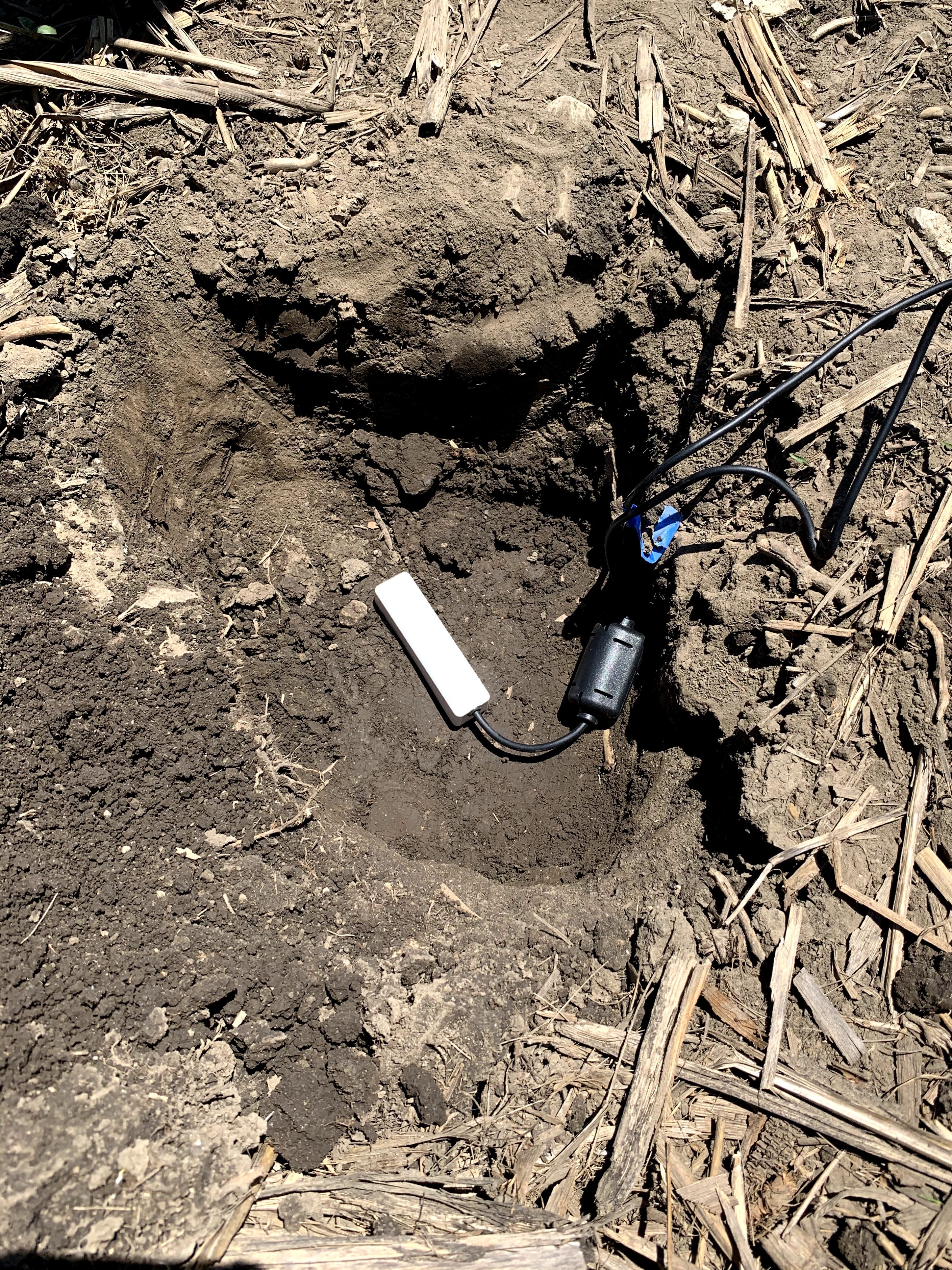}
    \caption{TEROS 12 SN placed at 15 cm depth.}
    \label{fig:soil-2}
    \end{minipage}
\end{figure}

\customsec{TEROS 12 Soil Sensor}
We used a commercial soil moisture sensor TEROS 12~\cite{teros},  as shown in (Fig.~\ref{fig:soil}; Fig.~\ref{fig:soil-2}). The TEROS 12 sensor for soil moisture, temperature, and electrical conductivity is reliable, easy-to-use, robust, yet economical. It incorporates METER's~\cite{meter} 70 MHz trademark circuitry with a durable epoxy filling and firmly attached, sharpened stainless steel needles that can slide quickly into the soil. The probe is resistant to salts, preventing sensor degradation. It features very low power consumption, and offers improved accuracy over a longer period of time with a high resolution.



\begin{figure}[htbp]
    \centering
    \begin{minipage}[t]{0.47\columnwidth}
    \centering
    \includegraphics[width=\textwidth]{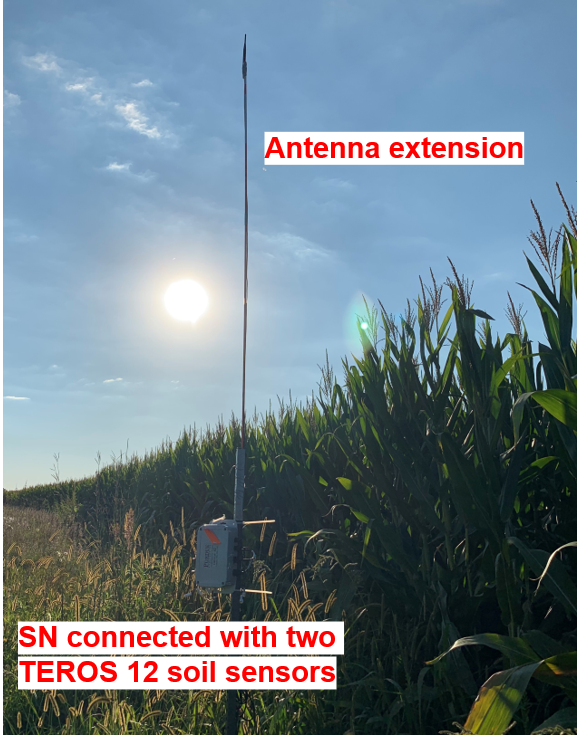}
    \caption{SN with antenna extension deployed during summer.}
    \label{fig:node-example}
    \end{minipage}
    \hspace{0.02\linewidth}
    \begin{minipage}[t]{0.47\columnwidth}
    \centering
    \includegraphics[width=\textwidth]{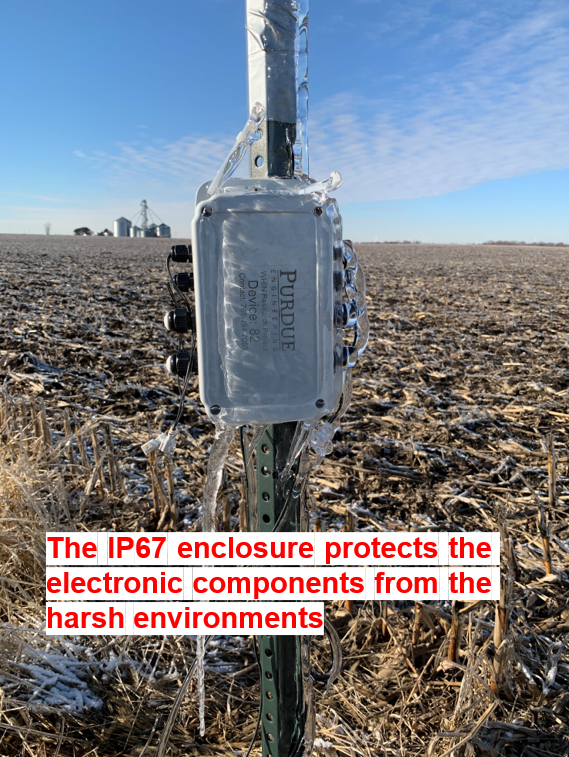}
    \caption{The IP67 enclosure providing protection in winter.}
    \label{fig:node-example-2}
    \end{minipage}
\end{figure}

\customsec{Enclosure}
The enclosure for the SN can achieve the IP67 Standard, which ruggedizing it. 
IP stands for ``Ingress Protection'' and is the International Protection Marking per IEC standard 60529. 
{The most common use of an IP rating is to show how protected a product is from a solid or liquid entering the product.}
The first digit indicates the level of protection a product has from a solid, \eg dust. The second digit indicates the level of protection from liquid intrusion in certain volumes, pressures, or temperatures. So, a sensor that is IP67 rated will remain protected and fully operational in most industrial applications, including those where it is exposed to water spray, rain, debris, etc. The ``6'' indicates the sensor is completely protected against solid objects from entering the sensor, including dust, while the ``7'' indicates the sensor can be completely submerged in 1 meter of water for up to 30 minutes before the moisture penetrates the housing.
An example of the deployed sensor is shown in Fig.~\ref{fig:node-example} and Fig.~\ref{fig:node-example-2} where the SN keeps working, even after being coated in ice in winter. 



\subsection{Gateway}
\label{subsec:gateway}
The Gateway has two main functions: collecting the data from SNs deployed in the field and uploading the data to the backend database server.
First, the Gateway collects the SN's measurements using LoRa modulation. Then, it backs up the data locally before transferring the readings to the backend over a TCP/IP connection.
In cases of low network connectivity, the Gateway uses a multi-hop approach. Here, the Gateway forwards the data to a nearby LoRaWAN Gateway---using the LoRaWAN protocol---transferring the readings to the backend through TCP/IP. We use this multi-hop setup in our LoRaWAN deployments, as shown in Fig.~\ref{fig:multiple-hop}.
We built our LoRa Gateway using a Raspberry Pi. The device is equipped with a LoStik transceiver that supports LoRa modulation for reception and transmission, as well as an LPWAN concentrator, compatible with the LoRaWAN protocol. The LoStik is used to collect the measurements from the SNs, and the LPWAN concentrator allows multi-hop routing using LoRaWAN specification.

\begin{figure}[htb]
    \centering
    \includegraphics[width=\columnwidth]{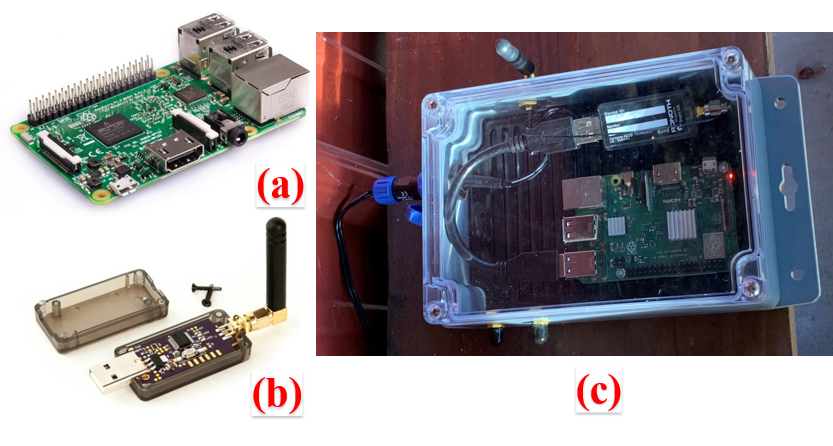}
    \caption{(a) RPi 4 Model B, (b) LoStik Module with USB port, (c) LoRa Gateway: RPi and LoStik in an IP65 enclosure.}
    \label{fig:gateway}
    \vspace{-2mm}
\end{figure}

\customsec{Raspberry Pi}
Raspberry Pis are lightweight and popular edge devices, varying in price from \$35 to \$140, and with an easily programmable interface for custom applications.
Raspberry Pi (RPi) 4 Model B, shown in Fig.~\ref{fig:gateway}(a), has a 1.5 GHz 64-bit quad core ARM Cortex-A72 processor, on-board 802.11ac WiFi, full gigabit Ethernet. The RPi 4 is powered via a USB-C port, enabling additional power to its downstream peripherals.
The operating system, RPi OS, is a Debian-based Linux distribution. The Gateway software runs on the RPi, which we manage remotely using SSH (Secure Shell) and VNC (Virtual Network Computing) protocols.

\customsec{LoStik}
The LoStik~\cite{lostik} is a USB LoRa device, as shown in Fig.~\ref{fig:gateway}(b). The LoStik is based on the RN2903/RN2483 LoRa Modules by Microchip, which enables seamless connectivity with LoRa/LoRaWAN compliant devices. The USB dongle is affixed into the Raspberry Pi to provide a single communication channel to receive and send LoRa packets.

\customsec{LPWAN Concentrator}
A LPWAN Concentrator module based on Semtech SX1302 enables integration of the RPi with other network equipment with LPWAN capabilities. The Concentrator can detect uninterrupted packets at 8 different spreading factors, providing 10 channels with continuous demodulation.

\subsection{Backend Server} 
\label{subsec:backend-server}
We support our backend on a server with the following specifications: 8 Intel(R) Xeon(R) W-2123 CPU cores at 3.60GHz, NVIDIA Quadro P400 with 256 Pascal CUDA cores and 2 GB of GPU memory, 32GB Memory, 3TB Disk, using Ubuntu 18.04.4 LTS. This server provides the following software components: a NoSQL database and a Web Server (frontend) for \emph{LoRa} deployments, and a time-series database and application server for \emph{LoRaWAN} deployments.

\customsec{Databases}
Specifically for \emph{LoRa}, we use MongoDB to store all the data collected from rural or urban areas.
MongoDB is a NoSQL database that organizes data collection in JSON-like documents. The dynamic nature of data collected on the field requires a schemaless database like MongoDB. 
In the meanwhile, for \emph{LoRaWAN}, we use InfluxDB, a time-series database.
These databases are specifically built to handle metrics and time-stamped events. Furthermore, the data is indexed by the SN's identifier, which is useful to query the measurements from the visualization layer. We plan to integrate our autotuning software for increasing the performance of these databases both in conventional~\cite{rafiki, sophia, optimuscloud} and serverless environments~\cite{mahgoub2021sonic}.

\customsec{Application Server}
The application server is part of the LoRaWAN stack. 
The server handles the LoRaWAN application layer, including handling join requests and session keys, uplink data decryption and payload decoding, downlink queuing, and data encryption. 
We use ChirpStack~\cite{chirpstack}, an open source solution, for our IoT network. ChirpStack provides a RESTful API and a Web interface to manage users, organizations, applications, and devices related to the LoRaWAN network. 
Chirpstack, for example, can inventory and manage all LoRaWAN SNs deployed, configure gateways connected to the IoT network, and monitor the SN condition.

\customsec{Web/Visualization Server}
We also built a website on a Web server using Django, a high-level Python Web framework. 
The website allows the user to query the measurements and visualize them on charts or download them in a CSV file for offline analysis.
A visualization layer for LoRaWAN deployments stored in the InfluxDB time-series database is also included in our IoT network.
We use Grafana~\cite{grafana}, an open-source platform for monitoring and observability, which allows the user to query and visualize the data in dynamic dashboards. Moreover, the user can define alerts' rules to monitor changes. 

\subsection{Three-layer IoT Network Architecture} 
\label{subsec:network}
In this section, we describe our three-layer IoT network architecture. As we can see in Fig.~\ref{fig:topology}, we use different technologies, including wireless communication and wired communication, to assemble the whole network. 
In the bottom layer, the SNs send data to the Gateway through either LoRa or BLE. Then, the Gateway uploads the collected data to the backend server using a secure connection over TCP/IP or LoRaWAN. Finally, the end-user can visualize, in real-time, the measurements collected from the field.
{Next, we describe the three layers of our architecture.}

\subsubsection{First Layer: Data Collection}
\label{subsubsec:data-collection}
When building our IoT network, we face multiple decisions regarding the communication protocol between SNs and Gateway and among the SNs themselves, \eg a mesh network proposed in~\cite{jiang2020hybrid}. If nodes are close to each other (1 to 100 meters), Wireless Local Area Network (WLAN) protocol, such as Bluetooth, WiFi, and ZigBee, are more suitable. In contrast, for long-range communications (a few kilometers in urban areas to over 10 km in rural areas), we need LPWAN protocols. Table~\ref{tab:protocol} presents an overview of some of the most relevant LPWAN available standards: LoRaWAN, Narrow Band IoT (NB-IoT), and Sigfox.

\begin{table*}[h!]
    \small
    \centering
    \caption{Specification of LPWAN protocols: LoRaWAN, NB-IoT, and Sigfox.}
    \begin{tabularx}{\textwidth}{|>{\raggedright\arraybackslash}X
    |>{\centering\arraybackslash}X
    |>{\centering\arraybackslash}X
    |>{\centering\arraybackslash}X|}
    \hline
    \textbf{Protocols} & \textbf{LoRaWAN} & \textbf{NB-IoT} & \textbf{Sigfox} \\ 
    \hline
    \hline
    Spectrum & Unlicensed ISM radio bands & Licensed LTE frequency bands & Unlicensed ISM radio bands \\
    \hline
    Bandwidth & 125 or 250kHz & 180kHz & 192Hz \\
    \hline
    Max data rate & 5.5kbps & 127kbps & 100bps \\
    \hline
    Max messages/day & 30 seconds air-time per device per day & Unlimited & Uplink: 140, Downlink: 4\\
    \hline
    Max payload length & 242 bytes & 1,600 bytes & 12 bytes\\
    \hline
    Range & Urban: 2-5 km, Rural: 10-15 km & Urban: 1 km, Rural: 10 km& Urban: 3-10 km, Rural: 30-40 km\\
    \hline
    Interference immunity & High & Low & High \\
    \hline
    Authentication \& encryption & AES 128 bit & 3GPP (128 to 256 bit) & N/A \\
    \hline
    Cost & Device (\$4-\$6), Own network infrastructure (\$400-\$1000), Network operation and maintenance & Device (\$6-\$12), SIM card (\$1-\$2), Subscription fees (\$5-\$30)& Device (\$5-\$10), Subscription fees (\$8-\$12)\\
    \hline
    \end{tabularx}
    \label{tab:protocol}
\end{table*}

\customsec{LoRaWAN}
It is one of the most promising LPWAN solutions and is currently gaining traction to support IoT applications and services based on the LoRa radio modulation technology. LoRa modulation adopts the frequency-shift keying (FSK) and Chirp Spread Spectrum (CSS) operating at 868 MHz (Europe) and 915 MHz (North America) ISM band, promising ranges upto 20-25 km. LoRa also supports an Adaptive Data Rate (ADR) from 300 bps to 50 kbps for a 125 kHz bandwidth and supports configurable chirp duration (spreading factor) to maximize both the battery life of each device and the overall capacity available through the system. LoRaWAN defines the MAC communication protocol and the system architecture for the network. It uses a hub-and-spoke topology in which every LoRa node can communicate directly with the Gateway module. 

\customsec{NB-IoT}
The 3rd Generation Partnership Project (3GPP) is a joint initiative of Asia, Europe, and North America telecommunications standardization organizations to produce global specifications for the Universal Mobile Telecommunication System (UMTS). 3GPP standardized a set of low cost and low complexity devices targeting Machine-Type-Communications (MTC) in Release 13. In particular, 3GPP addresses the IoT market using a threefold approach by standardizing the enhanced Machine Type Communications (eMTC), the NB-IoT (Narrowband IoT), and the EC-GSM-IoT. Specifically, NB-IoT can operate in a system bandwidth as narrow as 200 kHz and supports deployment in spectrum originally intended for GSM or LTE. It also supports a minimum device bandwidth of only 3.75 kHz. This design affords NB-IoT an energy efficient operation, while coexisting with existing GSM and LTE network, reducing deployment costs~\cite{narrowband2017}. 

\customsec{Sigfox} 
Sigfox is a proprietary Ultra Narrow Band (UNB) LPWAN solution that operates in the 869 MHz (Europe) and 915 MHz (North America) bands, with an extremely narrow bandwidth (100Hz). It is based on Random Frequency and Time Division Multiple Access (RFTDMA) and achieves a data rate of around 100 bps in the uplink, with a maximum packet payload of 12 Bytes. However, Sigfox limits the amount of data per device in the network to 14 packets/day. These limitations, together with a business model where Sigfox owns the network, have shifted the interest to LoRaWAN, which is considered more flexible and open.

As we can see, none of the previous LPWAN protocols is a perfect scheme. However, considering our application is a local ad-hoc network designed for digital agriculture, LoRa modulation presents a reasonable solution. Its low energy consumption, long-range, privacy features, and flexibility~\cite{devalal2018lora, sinha2017survey} fit our needs. 
Similar to the hybrid LPWAN proposed in~\cite{jiang2020hybrid}, we choose LoRa over LoRaWAN to support a broader coverage since the latter only supports a star
topology---SNs connected directly to the Gateway.  Nevertheless, in urban areas or widespread rural areas, a mesh network can extend the IoT network coverage. 
Fig.~\ref{fig:gateway}(c) shows a deployed Gateway consisting of a Raspberry Pi and LoStik in an IP56 enclosure~\cite{pi-package}, ruggedizing it. The configuration of the Gateway can be customized depending on the deployed area. For instance, the spreading factor can vary from 7 to 12, where a higher value denotes a broader coverage.

\subsubsection{Second Layer: Data Upload}
\label{subsubsec:data-upload}
After collecting data from SNs to the Gateway, we need to upload the respective measurements to the backend servers. 
{Considering that the Gateway may not be in the range of any network with Internet access (either through a wired or wireless connection), we have two possible scenarios. }
If the Gateway has Internet access, the Gateway uploads the measurements using a Transport Layer Security (TLS), or formerly, Secure Sockets Layer (SSL). In contrast, in the second scenario, the Gateway relies on LoRaWAN to forward the collected readings to a nearby Gateway with Internet access. 
As described earlier, LoRaWAN uses AES (Advanced Encryption Standard), with a symmetric key of 128 bits to encrypt the data.

\subsubsection{Third Layer: Data Persistence and Visualization}
\label{subsubsec:data-visualization}

\begin{figure*}[h!]
    \centering
    \begin{adjustbox}{varwidth=0.85\textwidth,margin=0 {\abovecaptionskip} 0 0, frame=0.25pt }
    \includegraphics[width=\textwidth]{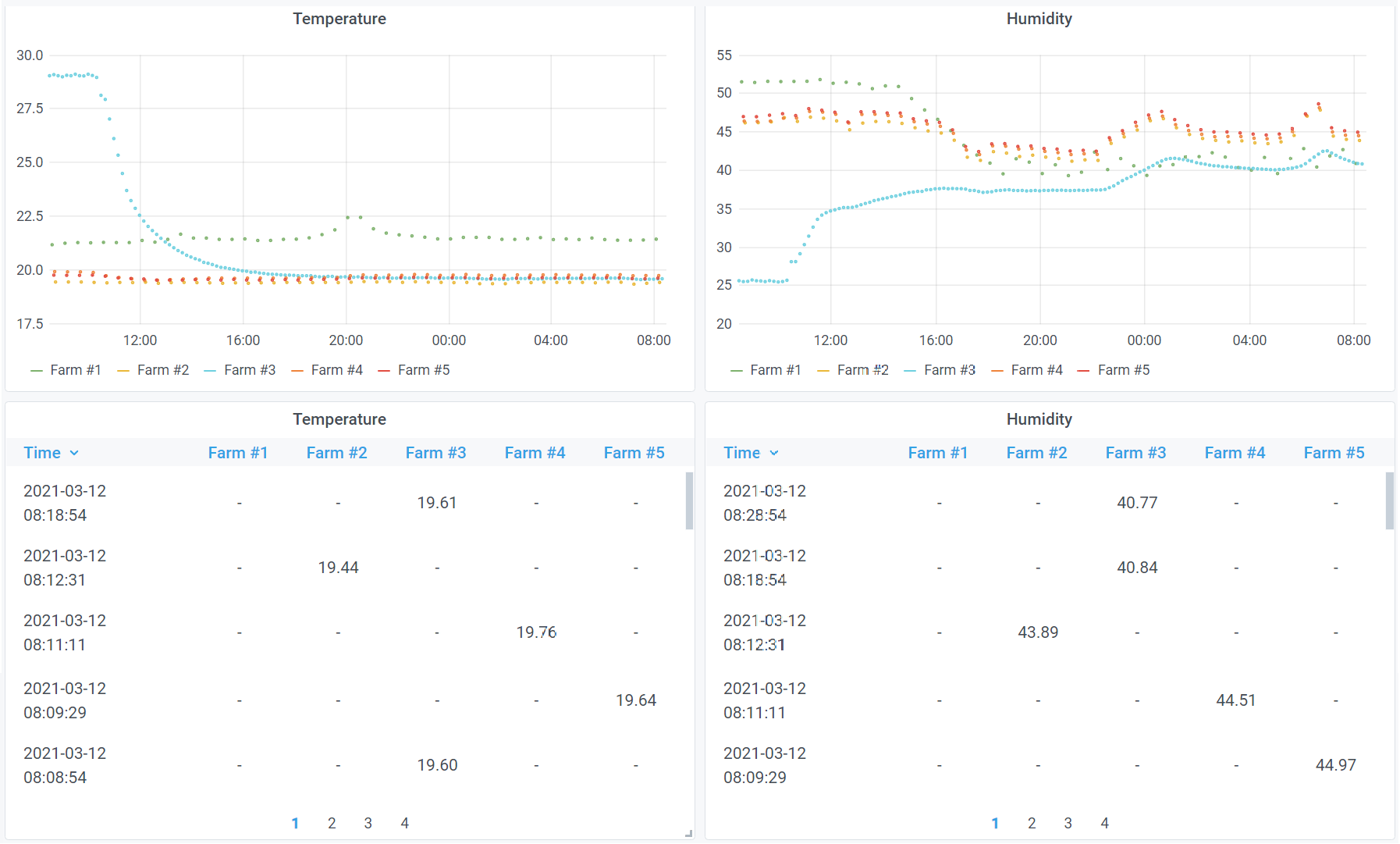}
    \end{adjustbox}
    \caption{Grafana dashboards including the temperature and humidity of 5 different SNs of our IoT network deployment.}
    \label{fig:grafana}
   \vspace{1em}
\end{figure*}

Our backend includes the following open-source servers: a NoSQL database (MongoDB), an application server (ChirpStack), a time-series database (influxDB), and a visualization server (Grafana). The NoSQL database stores the measurements uploaded through the network. The user can visualize these readings using a frontend website (Django framework). 
The frontend provides access to the data and allows to export it in a CSV format for further analysis outside of our platform.
On the other hand, the application server coordinates the communication with the LoRaWAN gateways deployed in the field. This server is integrated with a time-series database, influxDB, specially useful to store real-time data. Finally, the user can access the measurements using a Grafana server Fig.~\ref{fig:grafana}, which provides a flexible interface to query, visualize, and analyze the measurements.

\section{Emulated Testbed}
\label{sec:lab-testbed}




\begin{figure}[htb]
    \centering
    \includegraphics[width=\columnwidth]{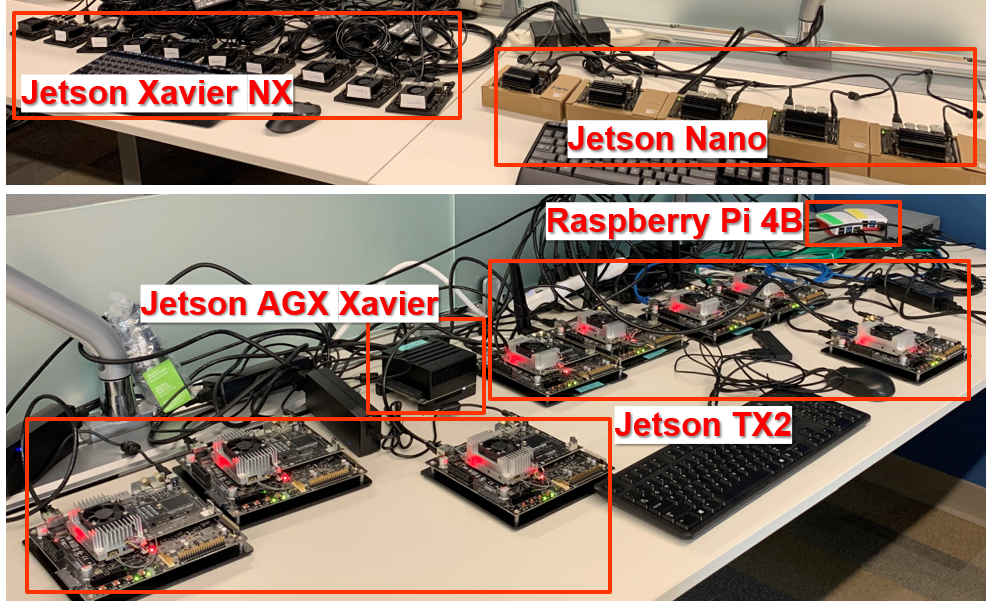}
    \caption{Heterogeneous testbed using NVIDIA Jetson devices and Raspberry Pis in a local area network (LAN).} 
    \vspace{-2mm}
    \label{fig:testbed}
\end{figure}
In addition to our living testbed, we also set up an emulated heterogeneous testbed in our lab using NVIDIA Jetson-class devices. The setup is shown in Fig.~\ref{fig:testbed}. 
Having an emulated testbed allows us to fine-tune the hardware-software interactions of the system in a controlled lab environment. Ironing out the kinks, stress testing, and proactively implementing additional layers of redundancy will result in a more resilient field deployment. 
Arguably, the most important benefit is that it allows us to fast-forward new feature implementation, major system changes, and prototyping.

\subsection{Devices}
\label{subsec:jetson-devices}
Table~\ref{tab:jetson} summarizes the specification of the NVIDIA devices in our emulated testbed. 
Besides, they could be deployed in the field or be carried on an aerial drone if we require a mobile high-performance SN. 

\begin{table*}[h!]
    \centering
    \small
    \caption{Comparison between Jetson modules: Nano, TX2, Xavier NX, and AGX Xavier.}
    \begin{tabularx}{\textwidth}{|>{\hsize=.6\hsize\raggedright\arraybackslash}X
    |>{\hsize=.68\hsize\centering\arraybackslash}X
    |>{\hsize=.68\hsize\centering\arraybackslash}X
    |>{\hsize=.68\hsize\centering\arraybackslash}X
    |>{\hsize=.68\hsize\centering\arraybackslash}X
    |>{\hsize=.68\hsize\centering\arraybackslash}X|}
    \hline
    \textbf{Models} & \textbf{Raspberry Pi 4B} & \textbf{Nano} & \textbf{TX2} & \textbf{Xavier NX} & \textbf{AGX Xavier}\\ 
    \hline
    \hline
    CPU & Broadcom BCM2711, Quad core Cortex-A72 (ARM v8) 64-bit SoC @ 1.5GHz & Quad-Core ARM Cortex-A57 MPCore @ 1.5GHz & Dual-Core NVIDIA Denver 2 64-Bit CPU and Quad-Core ARM Cortex-A57 MPCore processor @ 2.0GHz & 6-core NVIDIA Carmel ARMv8.2 64-bit CPU 6MB L2 + 4MB L3 @ 1.9GHz & 8-core NVIDIA Carmel Armv8.2 64-bit CPU 8MB L2 + 4MB L3 @ 2.3GHz\\
    \hline
    GPU & Broadcom VideoCore VI @ 500MHz & 128-core NVIDIA Maxwell GPU @ 921MHz & 256-core NVIDIA Pascal GPU @ 1300MHz & 384-core NVIDIA Volta GPU with 48 Tensor Cores @ 1100MHz & 512-core NVIDIA Volta GPU with 64 Tensor Cores @ 1377MHz\\
    \hline
    Memory & 2GB, 4G or 8GB 32-bit LPDDR4 @ 3200MHz - 25.6GB/s & 4 GB 64-bit LPDDR4 @ 1600MHz - 25.6GB/s & 8 GB 128-bit LPDDR4 @ 1866MHz - 59.7GB/s & 8 GB 128-bit LPDDR4x @ 1600MHz - 51.2GB/s & 32 GB 256-bit LPDDR4x @ 2133MHz - 136.5GB/s \\
    \hline
    Storage & MicroSD & 16 GB eMMC + MicroSD & 32 GB eMMC + MicroSD & 16 GB eMMC + MicroSD & 32 GB eMMC + MicroSD\\
    \hline
    Power modes & 15W & 5W/10W & 7.5W/15W & 10W/15W & 10W/15W/30W\\
    \hline
    DL Accelerator & - & - & - & 2x NVDLA Engines & 2x NVDLA Engines \\
    \hline
    AI Performance & 32 GFLOPS & 472 GFLOPS & 1.33 TFLOPS & 21 TOPS & 32 TOPS \\
    \hline
    Price & \$55 - \$100 & \$100 & \$400 & \$400 & \$700 \\
    \hline
    \end{tabularx}
    \label{tab:jetson}
\end{table*}

\noindent{\customsec{Jetson Nano}}
NVIDIA Jetson Nano is an embedded system-on-module (SoM) including an integrated 128-core Maxwell GPU, quad-core ARM A57 64-bit CPU, 4GB LPDDR4 memory, along with support for MIPI CSI-2 and PCIe Gen2 high-speed I/O. Jetson Nano runs Linux and provides 472 GFLOPS of FP16 compute performance with 5-10W of power consumption.
\noindent{\customsec{Jetson TX2}}
NVIDIA Jetson TX2 is an embedded SoM with dual-core NVIDIA Denver2 + quad-core ARM Cortex-A57, 8GB 128-bit LPDDR4 and integrated 256-core Pascal GPU. Jetson TX2 runs Linux and provides greater than 1TFLOPS of FP16 compute performance in less than 7.5 watts of power.
\noindent{\customsec{Jetson Xavier NX}}
NVIDIA Jetson Xavier NX is an embedded SoM including an integrated 384-core Volta GPU with Tensor Cores, dual Deep Learning Accelerators (DLAs), 6-core NVIDIA Carmel ARMv8.2 CPU, 8GB 128-bit LPDDR4x with 51.2GB/s of memory bandwidth, hardware video codecs, and high-speed I/O including PCIe Gen 3/4, 14 camera lanes of MIPI CSI-2. Jetson Xavier NX runs Linux and provides up to 21 TeraOPS (TOPS) of compute performance in user-configurable 10/15W power profiles.
\noindent{\customsec{Jetson AGX Xavier}}
NVIDIA Jetson AGX Xavier is an embedded SoM including an integrated Volta GPU with Tensor Cores, dual DLAs, octal-core NVIDIA Carmel ARMv8.2 CPU, 32GB 256-bit LPDDR4x with 137GB/s of memory bandwidth, and 650Gbps of high-speed I/O including PCIe Gen 4 and 16 camera lanes of MIPI CSI-2. Jetson AGX Xavier runs Linux and provides 32 TOPS of compute performance in user-configurable 10/15/30W power profiles.

\subsection{SLURM}
\label{subsec:slurm}
SLURM is an open-source distributed computing workload manager. SLURM allows multiple devices to work in parallel on a single task, as long as the application is written to execute in parallel mode. 
In the context of this paper, SLURM acts as a resource scheduler for the emulated testbed. We pre-train ML models on more powerful devices at the edge or cloud. 
Then SLURM could allow us to run inference on embedded devices in the field, resulting in system with significantly less data transferring.
Traditionally, SLURM is most prevalent in x86 architecture builds, and only recently began gaining traction in the ARM world. With significant numbers of ARM SoCs being deployed, we expect that ARM-based distributed computing will see significant adoption.
SLURM is used to manage ML training tasks on distributed GPU clusters~\cite{awan2019scalable, campos2017distributed}, parallelizing the training process. 
In our testbed, we leverage SLURM to distribute the inference task on multiple heterogeneous Jetson devices to accelerate the prediction process. Main challenge with this is finding the sweet spot of task distribution across multiple devices while keeping resource management, storage I/O, and power consumption overhead to a minimum. 
Another benefit of distributing a task across multiple devices is hardware redundancy, increasing the system's reliability. For example, if we rely on a single device to do the object detection task, a hardware failure will kill the system. However, if we spread the task across multiple devices, the system will only fail if all of the devices concurrently go offline. 
Finally, idle standby time can be leveraged to divert workloads to idle devices for resource efficiency.

\subsection{Tegrastats}
\label{subsec:tegrastats}

\begin{figure}[t!]
    \centering
    \includegraphics[width=\columnwidth]{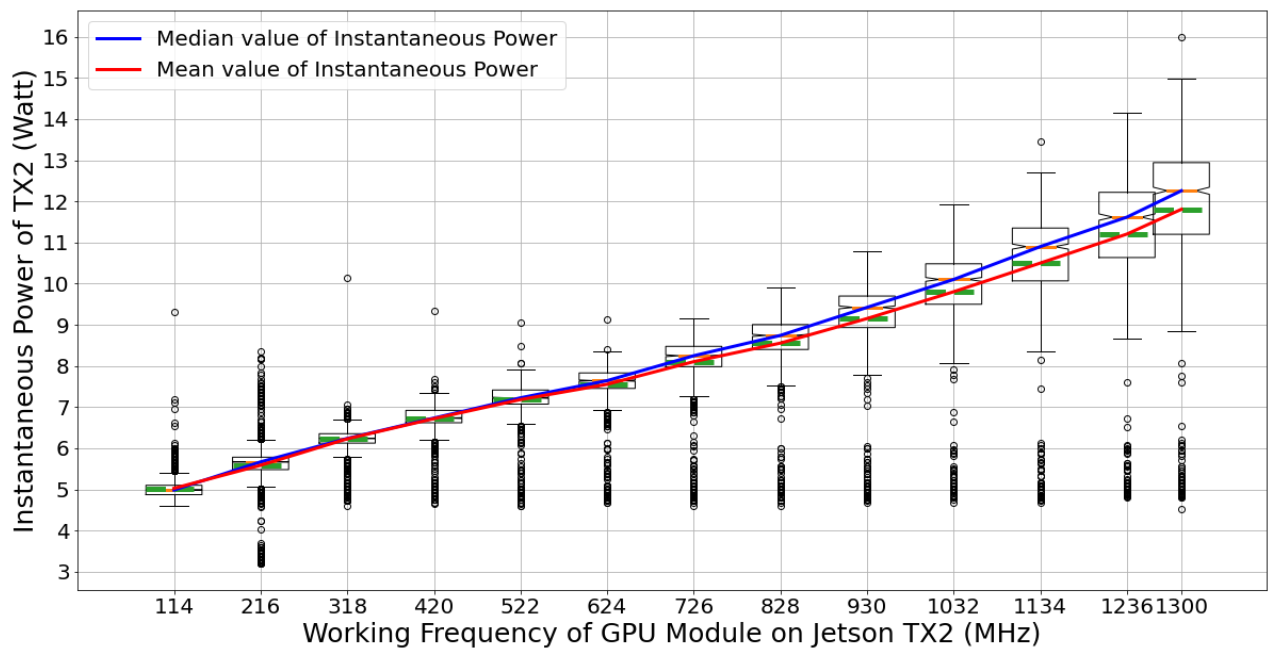}
    \caption{Jetson TX2's instantaneous power consumption vs DVFS setting of the GPU module while profiling ApproxDet's~\cite{xu2020approxdet} 40 approximation branches. The current power consumption of the board is directly influenced by the DVFS setting of the GPU module while running object detection tasks.}
    \label{fig:energy}
    \vspace{-5mm}
\end{figure}

The Jetson series products provide the {\em tegrastats} utility~\cite{tegrastats}, which reports important memory, processor, and disk usage information. This utility can be easily accessed by running the command ``sudo tegrastats.'' The detailed information includes RAM usage, CPU utilization for each core, GPU utilization, power consumption for main hardware modules, etc. 
The output can be observed in real time, or piped to a file for future analysis. In our experiments, we use the information from {\em tegrastats} to analyze the relationship between power consumption, and the intensity of computation tasks. We also use it to observe the change of CPU and GPU utilization when specific operations are performed. Fig.~\ref{fig:energy} shows an example of our profiling experiments of the current power consumption of ApproxDet's 40 execution branches~\cite{xu2020approxdet}. ApproxDet performs {\em approximate} video object detection, in a way that adapts to changes in content characteristics and changes in contention levels. This experiment is done on ImageNet Large Scale Visual Recognition Challenge 2015 validation dataset
when we use different Dynamic voltage and frequency scaling (DVFS) settings for the GPU module. The values for DVFS settings of the GPU module are provided by NVIDIA in the NVIDIA Jetson Linux Developer Guide~\cite{nvidia-doc}. 
{The instantaneous power information is collected by using {\em tegrastats}.}
As seen from the figure, with increasing frequency setting of the GPU module, the mean and median values of current power of the TX2 board keep increasing while running ApproxDet's execution branches. Another observation is that the variance of the energy consumption also increases as GPU frequency increases. That is because with a higher GPU frequency, a greater number branches can run (including the heavyweight branches, which may not have been possible at lower frequencies), resulting in greater variability because of the higher spread of the (possible) approximation branches. 
Further, we can tune the DVFS settings of CPU, GPU, and memory modules separately based on the API that the Jetson platform provides. 
This means the user can configure each module independently based on the application requirements (accuracy, latency, and energy).  

\begin{figure}[htbp]
    \centering
    \includegraphics[width=\columnwidth]{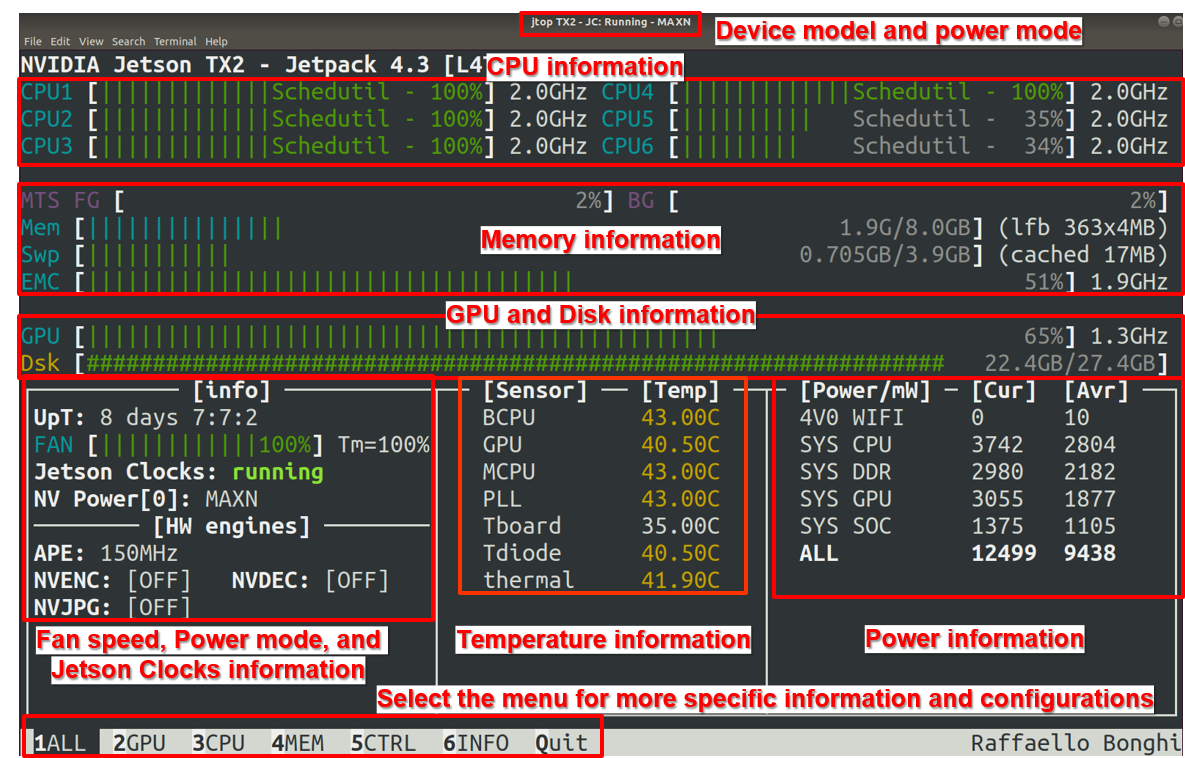}
    \caption{Jetson stats' GUI shows detailed working status information for each hardware module on the device.}
    \label{fig:jtop}
   \vspace{-2em}
\end{figure}

Based on the {\em tegrastats} utility, an open-source tool named {\em Jetson stats}~\cite{jtop} 
has been developed. It is a package for monitoring and controlling NVIDIA Jetson devices, showing various component-level performance statistics. GUI variant allows tweaking of various hardware settings. Fig.~\ref{fig:jtop} is an example of running {\em Jetson stats} on one of TX2 boards in our testbed. By observing the displayed values, we can easily observe the utilization levels of CPU, GPU, and memory. For example, if a GPU-intensive memory task is launched, the GPU frequency and utilization will become high. The temperature and power increases can also be observed. 


\subsection{Drones}
\label{subsec:drones}
Our embedded testbed is augmented with aerial drones~\cite{dronet21}. These complement the ground nodes in two ways. First, the drones periodically fly over the covered regions, and collect data that may have been missed due to communication holes. This interval can range from hourly to once a day. Drone-based data ferrying can help in areas with limited, or no wireless connectivity. This can happen due to physical conditions, such as growth of corn stalks, resulting in a loss of communication of nodes in the network chain. 
There is a protocol programmed into the SNs 
to send buffered data to the drones when drones indicate being within communication range. This form of data ferrying is well known in the literature~\cite{pu2019route} and works in our context as there can be short to long periods of network disconnection for some of the nodes. A second enabling factor is that much of the data also does {\em not} have strict latency requirements and can afford periodic visits by the drones. 
 Another use case is an on-demand summoning of a drone by the SNs. An SN has a limited computational capacity and can summon a drone when it senses that some event is unfolding that needs heavyweight processing such as streaming video analytics, object localization, object detection, or face recognition, resulting the actuated ``descent'' of the drone for higher precision. An SN can, through the regular (low bandwidth) LoRa channel, send a request to the backend server to dispatch a single, or multiple drones. The drone with a GPU-enabled embedded board can reach the SN, and perform the heavyweight analytics task for the time for which the event is happening. The raw video data does not need to be communicated over long ranges, but only over a short-range high bandwidth channel like BLE. Further, we have done preliminary work to optimize the drone coverage with a swarm of drones~\cite{dronet21}. We allow the drone to change its height to optimize the precision versus coverage when a swarm of drones is deployed in a farm (for on-demand data ferrying from the ground sensors). Overall, we have observed that the precision of the computer vision workload is a function of the drone height, rather than the drone speed. 

\section{Implementation and Evaluation}
\label{sec:implementation}
This section describes the techniques and challenges of the deployment of \name in production farms in northern and central Indiana and on Purdue campus.
\subsection{Deployment}
\label{subsec:deployment}
\begin{figure}[htb]
    \centering
    \includegraphics[width=\columnwidth]{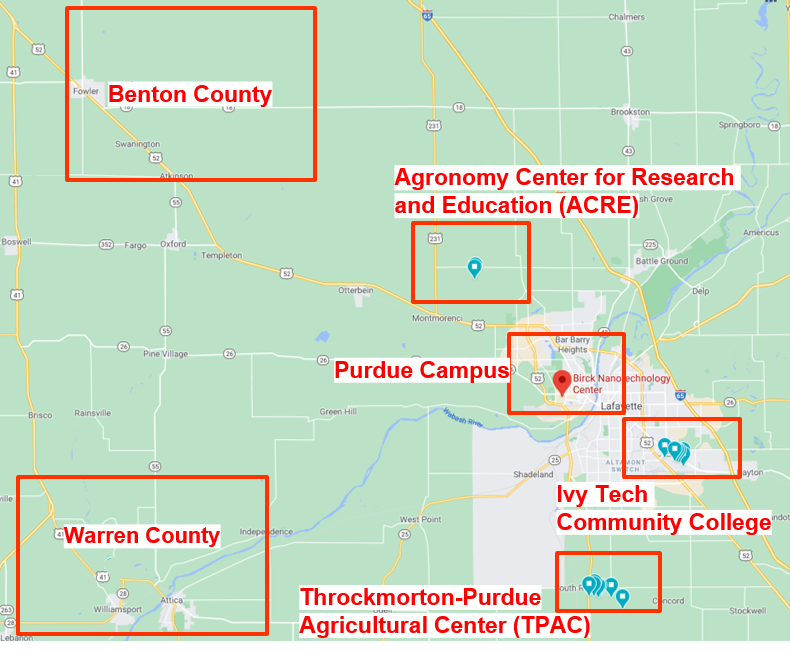}
    \caption{\name's IoT deployment map in Northern and Central Indiana: Purdue Main Campus, Purdue ACRE, Ivy Tech Community College, Throckmorton-Purdue Agricultural Center (TPAC), and commercial farms in Benton and Warren Counties.}
    \label{fig:map_deployed}
\end{figure}
As can be observed in Fig.~\ref{fig:map_deployed}, we deployed a living testbed of IoT nodes in urban areas (Purdue University main campus, West Lafayette) and rural areas (Purdue's experimental farms and commercial farms) across Indiana. 
\begin{figure}[htb]
    \centering
    \includegraphics[width=\columnwidth]{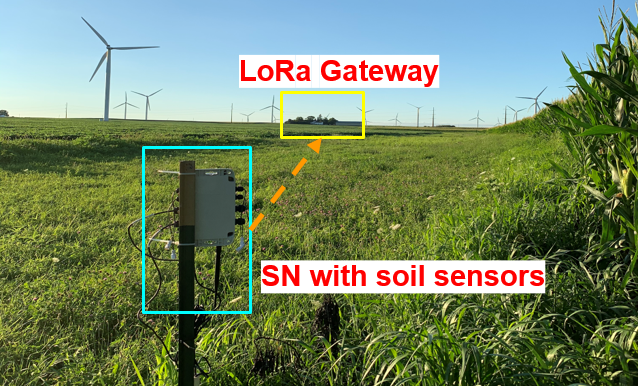}
    \caption{Deployment at a farm in Benton County: The SN is equipped with two TEROS 12 sensors and sends measurements to the LoRa Gateway inside the farmer's house via LoRa.}
    \label{fig:field_deployment}
\end{figure}
Shown in Fig.~\ref{fig:field_deployment}, an SN is set beside the cornfields, while the LoRa Gateway is located inside a house with a clear line-of-sight (LoS) propagation from the node. The SN is equipped with two TEROS 12 soil sensors, buried underneath the soil, to detect changes in the volumetric water content (VWC) at 12 and 6 inches depth, respectively. Each SN is capable of collecting the temperature and the relative humidity level using the on-board HDC2010 sensor included in the nodes, as described in Section~\ref{subsec:sensor-node}. 

SNs transmit the measurements periodically to the nearest Gateway using the LoRa communication protocol. 
Gateway receives the measurements and forwards them to the central database servers, as described in Section~\ref{subsubsec:data-upload}. 
As a contingency, the Gateway also stores the measurements locally in case of network failure.
\begin{figure}[htbp]
    \centering
    \includegraphics[width=\columnwidth]{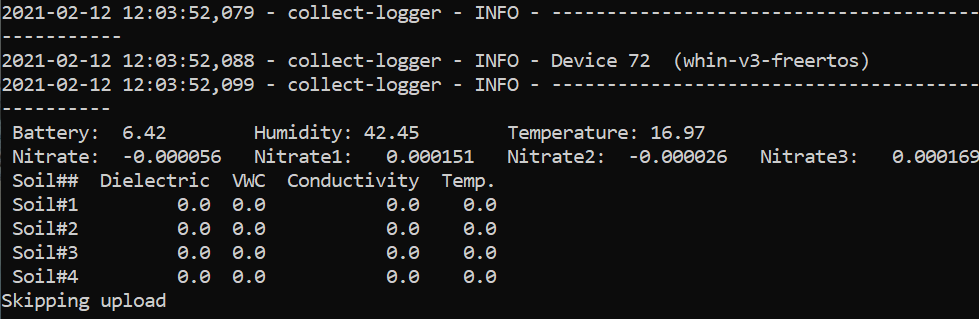}
    \caption{Readings on the Gateway: SN ID, firmware version, time stamp for received message, battery level, temperature, humidity, soil nitrate level, soil dielectric, conductivity, temperature, VWC.}
    \label{fig:receiver_output}
\end{figure}

\begin{figure*}[htbp]
    \centering
    \includegraphics[width=0.85\textwidth]{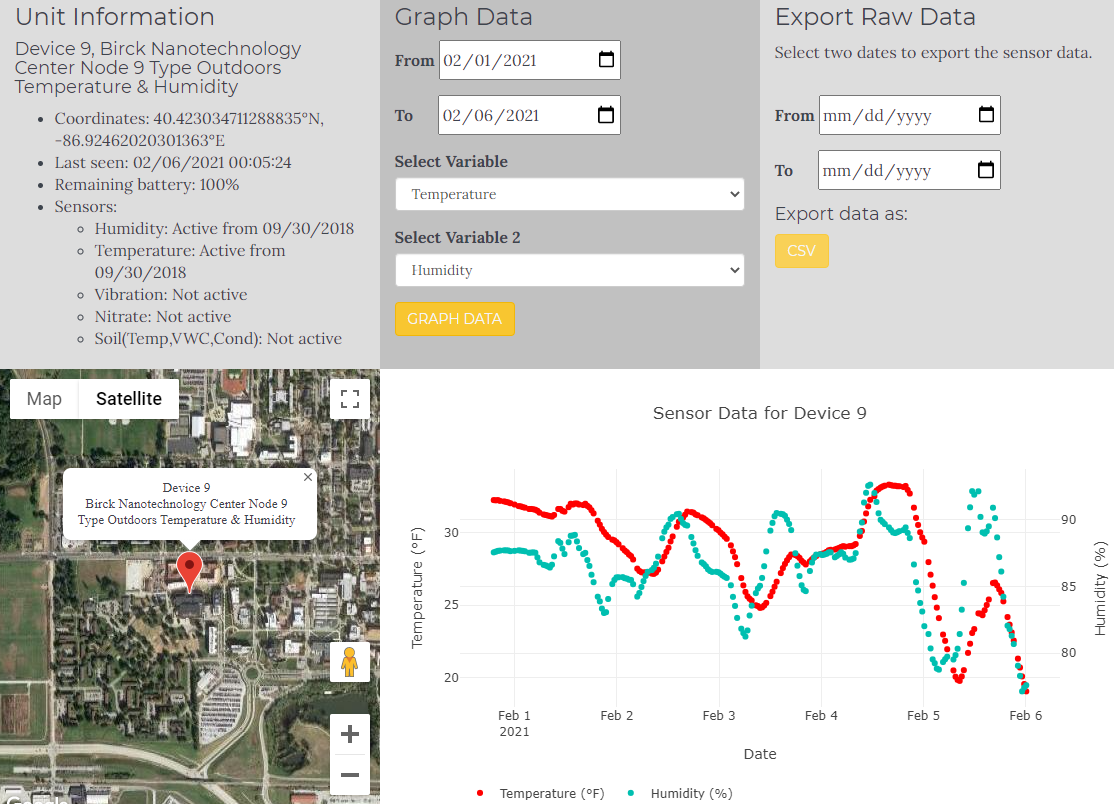}
    \caption{Website for SN readings, which include information such as SN ID, Location, GPS coordinates, last seen time, and the battery level. The website also allows to directly plot the data online or download it in .csv format.}
    \label{fig:website}
    \vspace{-1em}
\end{figure*}


Fig.~\ref{fig:receiver_output} shows an example of the output generated on a 
LoRa Gateway each time it receives the measurements. Primary information is received date and SN ID. Secondary information includes specific measurements for each metric: humidity, temperature, nitrate level, soil dielectric permittivity, conductivity, and VWC.
Once data is stored in the database server, it can be visualized from the frontend, as can be seen in Fig.~\ref{fig:website}. On the website, a user can see detailed location for the SN with its GPS coordinates, the last-seen time---a timestamp for the latest message, and the sensor installation information (\eg which sensor ports have been activated and when). We also provide the functionality \textit{Graph Data} for a user to plot the sensor measurements by choosing the date range and the sensor to use; \eg in Fig.~\ref{fig:website}, we plot the temperature and humidity data. Finally, user can also download the raw data within a date range for further analysis using the \textit{Export Raw Data} function.



\begin{figure}[htbp]
    \vspace{-3mm}
    \centering
    \includegraphics[width=\columnwidth]{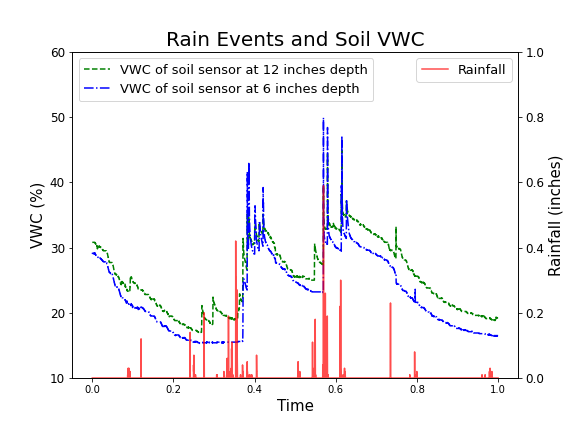}
    \caption{{Soil VWC (\%) at different depths (12'' and 6'', respectively) and rain events (inches) in the nearby location.}}
    \label{fig:soil-rain}
\end{figure}

{In Fig.~\ref{fig:soil-rain}, we show an example of combined analysis using our collected soil VWC data and the open source data from WHIN weather stations on rain events. Both data were collected from June 1, 2021 to August 13, 2021. Our soil sensor data has an interval of 30 minutes and the rain events' data from WHIN's weather station has an interval of 15 minutes. The WHIN weather station is 5.8 miles away from our soil sensor node in a rural area, which we believe, share the same weather patterns, including temperature, rainfall, etc. The X-axis represents the time, synchronized across the two locations. 
The left Y-axis is VWC (Volumetric Water Content) of soil sensors at two different depths (12 inches and 6 inches). The right Y-axis is the cumulative rainfall in the last 15 minutes. There are some insights we get from the figure:
\vspace{-2mm}
\begin{enumerate}[leftmargin=1em]
    \item As expected, the soil VWC measurements always follow the changes in rainfall in direction and magnitude. 
    \item The soil at greater depths has higher VWC. 
\end{enumerate}
}


\subsection{IoT Network Monitoring and Troubleshooting}
\label{subsec:network-monitoring}

\noindent{\customsec{Data Filtering}}
Data collected in the field can become corrupted before it gets stored in the backend database. This can happen either due to malfunction of device sensors, or signal interference during wireless communication. Our IoT network uses different filters at both, edge and cloud, to achieve end-to-end data integrity. These filters prevent anomalous data from reaching the front end which is the Data Visualization layer of {\name}, as shown in Fig.~\ref{fig:website}.
We use in-lab experiments to determine the threshold for each metric. For example, humidity measurement's reasonable range is between 0\% to 100\%; soil VWC's reasonable range is between 0 to 0.7, based on our experiment calculations.

\begin{figure}[htbp]
    \centering
    \includegraphics[width=\columnwidth]{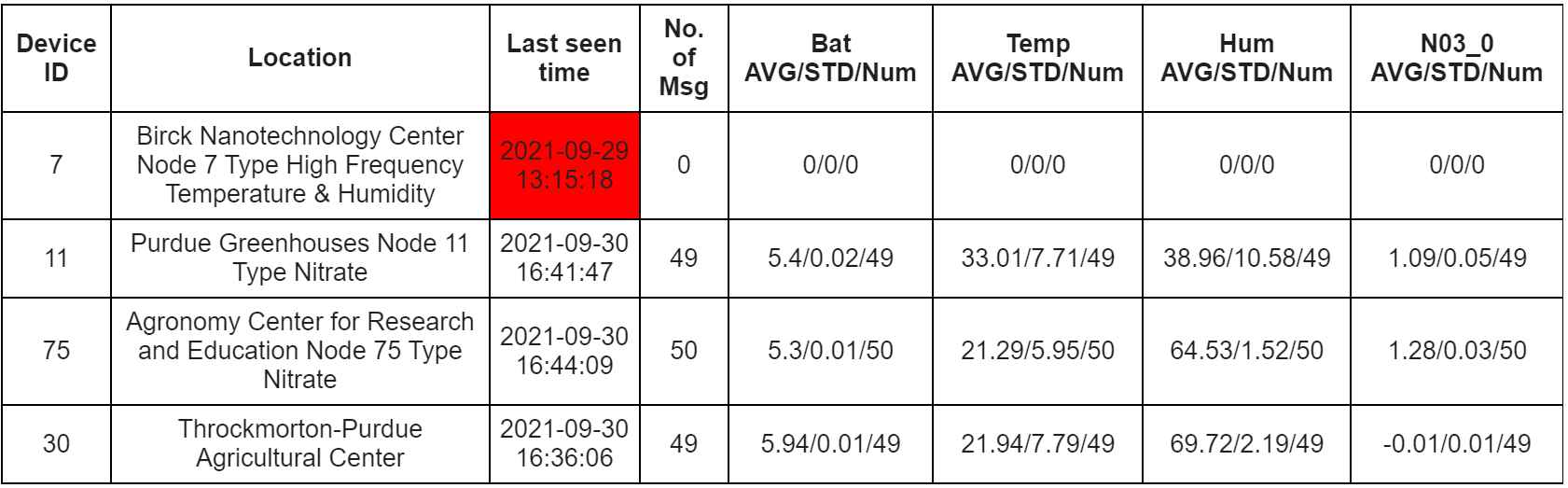}
    \caption{Email notification format from Network Monitoring and Anomaly Detection (NMAD), including: SN ID, location, last-seen time, number of messages received in a single day, mean and standard deviation of the measurements collected (\eg nitrate level, temperature, battery level). The red cell indicates that there is no recorded reading from the SN in the last 24 hours.}
    \label{fig:email}
    \vspace{-2mm}
\end{figure}

\customsec{Network Monitoring and Anomaly Detection}
{Production networks require constant monitoring and anomaly detection in several layers of the system. Network traffic monitoring and data integrity checks ensure long uptime and maintenance-free operation. Data anomalies can occur from natural causes (\eg data drift) and not necessarily from a security breach. Out of necessity, and as a part of \name, we developed NMAD, a server daemon that checks the status of the SNs deployed in the live testbed and sends reports or notifications to the system managers. Fig.~\ref{fig:email} shows an example of the email notifications. 
The notifications, whose frequency can be customized   (\eg hourly, daily), include the mean and standard deviation for every measurement. Moreover, the report highlights possible anomalies in the SNs. As an example, NMAD flags as anomaly any SN that has not communicated any reading in the last 24 hours.}
{We are including additional features in the NMAD, such as (near-real time) anomaly detection for new data based on the patterns from the existing data.}




\begin{figure}[htbp]
    \centering
    \includegraphics[width=\columnwidth]{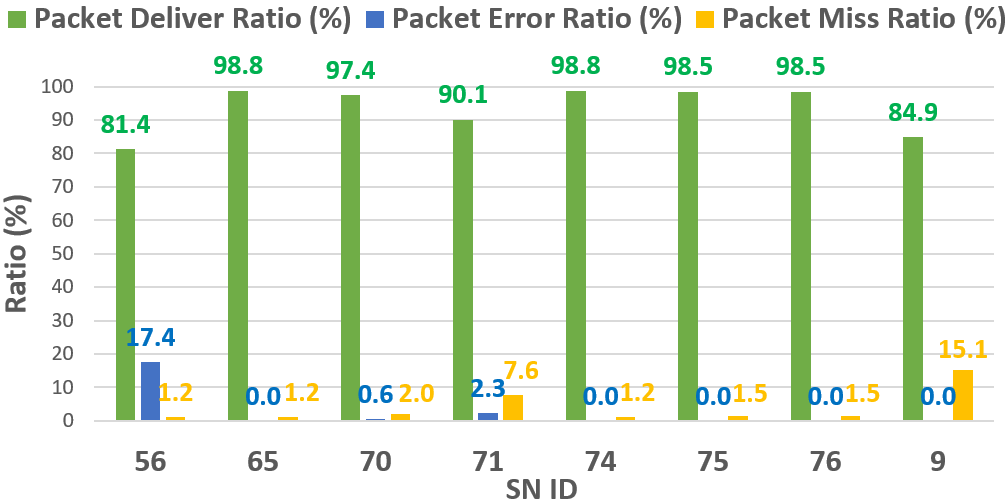}
    \caption{Packet Delivery Ratio (PDR), Packet Error Ratio (PER), and Packet Miss Ratio (PMR) over 7 days in our living testbed.}
    \label{fig:pdr}
\end{figure}


\subsection{Network Stability of Large-Scale Deployment}
\label{subsec:network-stability}
{We evaluate \name in terms of the stability of the network based on the following metrics: (a) Packet Delivery Ratio (PDR); (b) Packet Error Ratio (PER); and (c) Packet Miss Ratio (PMR).}
We report on network measurements from a sample selection of nodes from our living testbed. This includes measurements from farms (ACRE, Ivy Tech Community College, Warren County, and Benton County) and urban areas (Purdue Main Campus). During one week, February 23, 2021 to March 2, 2021, we analyzed the readings collected from these SNs in terms of PDR, PER, and PMR.
We calculate PDR, PMR, and PER as follows:
 
\begin{equation}
\small
\label{eq:pdr}
  PDR = \frac{\#\ of\ Received\ Normal\ Messages}{Expected\ \#\ of\ Messages}
\end{equation}

\begin{equation}
\small
\label{eq:per}
  PER = \frac{\#\ of\ Received\ Error\ Messages}{Expected\ \#\ of\ Messages}
\end{equation}

\begin{equation}
\small
\label{eq:pmr}
  PMR = \frac{\#\ of\ Missing\ Messages}{Expected\ \#\ of\ Messages}
\end{equation}


Where \begin{small}$PDR + PMR + PER = 1.0$\end{small}. Fig.~\ref{fig:pdr} summarizes the network stability profile. 
The communication interval for each SN is 30 minutes.
We use this interval to calculate the anticipated number of received messages in one week as: 7 * 24 * 60 / 30 $\approx$ 336.

As can be seen from Fig.~\ref{fig:pdr}, SN 56 has a higher PER than the rest of SNs. This is 
because of an anomaly where the post with the sensor had fallen to the ground, directly measuring a higher moisture level. SN \#9 is deployed in an urban area with high building density, with a resultant high PMR. With respect SN \#71, the slightly high PER and PMR is because there is a slope in the terrain between the SN and the gateway node, affecting the data transmission. The data collected by our living testbed is available at~\cite{purduewhin-sensor-list}.



\subsection{Network Stability of Two-Hop Deployment in Urban Area}
\begin{figure}[htbp]
    \centering
    \includegraphics[width=0.9\columnwidth]{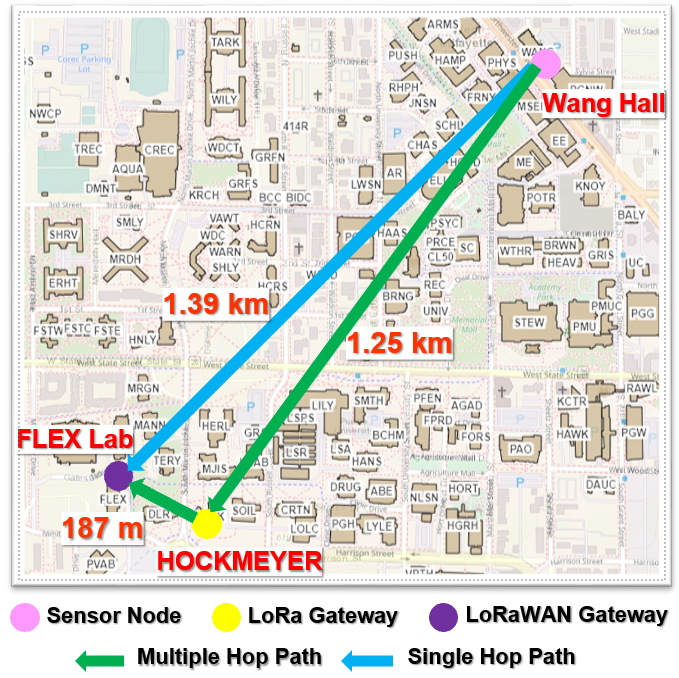}
    \caption{Purdue campus multiple-hop network map.}
    \label{fig:multiple-hop}
\end{figure}

{Because a building on the path can obstruct the signal, LoS propagation is critical when deploying LoRa devices (usually low-power wireless devices) in urban environments.
In addition, the signal will be attenuated with the reciprocal square of distance~\cite{mahapatra2016experimental}.
In practice, however, the LoS wireless propagation path is not always possible, especially in built urban environments. In order to establish an LoS wireless connection in a city, a Gateway or the nodes have to be deployed at heights, such as a building's roof or on a tall utility tower. Because of these constraints, the LoRa network's real communication range is often substantially lower than the theoretical value (5 km in urban area vs 20 km in rural area).
Our testbed provides a multiple-hop network to tackle this difficulty. 
Through a LoRa wireless connection, an SN transmits messages to a LoRa Gateway, which then forwards the messages to the LoRaWAN Gateway. We employ hybrid protocols because using LoRa for the first hop allows us to customize our local network. By using LoRaWAN for the second hop connection, we will be able to leverage the previously installed LoRaWAN Gateway and benefit from the standardization of the LoRaWAN protocol.
The signal is captured and amplified in the middle point---the LoRa Gateway---enabling us to communicate the data in a reliable manner to the LoRaWAN Gateway.}
{Fig.~\ref{fig:multiple-hop} shows the deployment map of our experiments on the Purdue campus.
We would like to send the signal from Wang Hall (about 21 meters above the ground) to Flex Lab (about 7 meters above the ground). However, a signal path wireless connection does not work because the buildings between them block the signal. By using our multi-hop network, first, the node sends the signal to the LoRa Gateway, located at the Hockmeyer building (about 13 meters above the ground). The LoRa Gateway collects the data, stores it locally, and then forwards it to Flex Lab by using LoRaWAN modulation. 
To evaluate the network quality, we measure the PDR, PER, and PMR for the two individual links, and then, the end-to-end connection. Results can be seen in Table~\ref{tab:multiple-hop}. 
\begin{table}[h!]
    \centering
    \caption{Multiple Hop Network Stability.}
    \begin{tabular}{|p{2.8cm}|p{1cm}|p{1cm}|p{1cm}|}
    \hline
    \textbf{Steps} & \textbf{PDR} & \textbf{PER} & \textbf{PMR} \\ 
    \hline\hline
    Node to LoRa Gateway & 62.15\% & 7.64\% & 30.21\% \\
    \hline
    LoRa Gateway to LoRaWAN Gateway & 88.27\% & 8.94\%  & 2.79\% \\
    \hline
    End-to-end & 54.86\% & 5.56\% & 39.58\% \\
    \hline
    \end{tabular}
    \label{tab:multiple-hop}
\end{table}
The first hop achieves a relatively low PDR and high PMR compared to the second hop because the propagation path for the first hop is not a perfect LoS propagation; its distance (1.25km) is much larger than the second hop (187m). The second hop is a clear LoS propagation, resulting in a higher PDR. In a real-world urban area, our multiple-hop network achieves over 55\% PDR, which a typical LoRaWAN single-hop network cannot.}

\section{Big Data Frameworks to the Rescue}
\label{sec:big-data}
\noindent Here, we will discuss the popular frameworks through which we will invoke the ML algorithms in our living testbed. This will involve open-source frameworks such as Apache Spark, streaming data processing frameworks such as Apache Flink, and techniques for distributing the ML processing among nodes in a cluster, both on-premise~\cite{sophia, rafiki} and in a conventional cloud~\cite{optimuscloud} or serverless environment~\cite{mahgoub2021sonic}, such as for video analytics workloads~\cite{xu2019approxnet, xu2020approxdet}. Serverless computing has become a promising cloud model where cloud providers run the servers and manage all administrative tasks (\eg scaling, capacity planning, etc.), while users focus on the application logic~\cite{schleier2021serverless}.
Due to its elasticity and ease-of-use, serverless computing is becoming increasingly popular for advanced streaming workflows such as data processing pipelines~\cite{pu2019shuffling}, ML pipelines~\cite{muller2020lambada}, and video analytics~\cite{fouladi2017encoding, ao2018sprocket, xu2020approxdet}. Major cloud providers have recently introduced serverless {\em workflow} services such as AWS Step Functions, Azure Durable Functions, and Google Cloud Composer, where applications are composed of a sequence of execution stages, which can be represented as a DAG~\cite{pu2019shuffling,elgamal2018costless}. DAG nodes
correspond to serverless functions ($\lambda$s) and edges represent the flow of data between dependent $\lambda$s, as designed in our system, SONIC~\cite{mahgoub2021sonic}. The serverless abstraction is attractive in this domain because 
the computation is often \textit{event driven}, \eg when a particular sensor is triggered, say to detect the sound of an oncoming combine. Thus, compute nodes \textit{do not} need to be active all the time but can be ``turned on" at the right time  to process an event. 
The system owner (who is typically not a computer scientist) 
is not responsible for making subtle resource provisioning decisions. A Reinforcement Learning (RL)-based solution can be attached to the serverless execution to make decisions such as how much memory to allocate to any container in the serverless execution. We have productively used such RL-based orchestration for controlling physical operations (in a manufacturing floor setting) in prior work~\cite{thomas2018minerva}. Another reason serverless appears attractive in this domain is that the owner is only charged based on the number of events processed, rather than for a length of time. This should make it economically attractive for this domain where event-driven processing is the norm. 

This infrastructure is in preparation for deployment of SNs within a farm and then connecting SNs across farms to form a federated setting. In the federated setting, with the privacy transforms, data acquisition and analysis can be done while maintaining anonymity using well-known techniques such as secure multi-party computation (MPC), especially their efficient variants~\cite{damgaard2013practical, ben2008fairplaymp}. 
Such techniques will guarantee that individual bad actors cannot break the confidentiality properties of our solution. Rather, only if greater than a threshold of the participants become bad actors will the system become vulnerable. This is feasible because today there exist ``push-button" solutions for MPC and our domain for many scenarios does not need very low latency (the baseline solutions in MPC are known to be slow). 

\noindent{\customsec{Streaming data processing}}
Since most of the data obtained in digital agriculture is real time, stream processing is preferred over batch processing. This adds several constraints: {\em first}, the data analytics code has to function at a rate at least as fast as the rate at which data is being generated; {\em second}, it has to calculate statistics (such as, range for normalization of data) without access to the entire data and based on some look-ahead window based on the workload dynamism (for example, for a more dynamic workload, the look-ahead window will be lower for higher accuracy, plus the algorithm should be configured for some degree of error handling); and {\em third}, the code has to have the right input-output interfaces so that it can ingest streaming data and output its results in a stream. Now we consider some popular open source streaming analytics frameworks---Apache Spark Streaming (\url{https://spark.apache.org/streaming/}), Apache Storm (\url{http://storm.apache.org/}), and Apache Flink (\url{https://flink.apache.org/}). These differ in the ways in which they can transform the data stream, \ie the kinds of operators that they support, the latency of processing, the programming languages they support, etc. 
\cite{carbone2015apache} provides an open-source stream processing framework. Flink, however, does not provide its own  data storage and needs to be supplemented by frameworks that do, \eg Kafka~\cite{garg2013apache} or Cassandra~\cite{cassandra2014apache}. Apache Spark~\cite{zaharia2016apache} is an alternative that can also be used for data parallelism. 

\noindent{\customsec{Batch data processing} }
There are some data analytics applications that need batch processing in this domain. This includes analytics that will be processed for strategic decision-making, without any real-time requirement. Batch data processing is done through data warehousing tools~\cite{chaudhuri1997overview, nargesian2019data} and analytics frameworks like Spark (as opposed to Spark-Streaming) that can ingest data from such warehouses. The ease of this mode of processing is that the analytics code can access the entire data in one shot. The challenge with this mode of processing is the large volumes of data. To fit within the resources of the compute nodes, the data has to be segmented and the analytics code then runs on the segmented data. 

\section{Conclusion and Future Work}
\label{sec:conclusion}
We have deployed three models for testing and refining \name. The \textit{first model} consists of a living lab with real, ruggedized sensors deployed in fields and networking innovations for low end-to-end latency. This is for more latency-sensitive tasks, such as near-live actuation of sensors on applicator for farm supplements and site-specific variable rate treatments. Further, there are multiple-hop networks connecting the sensor nodes on the farm and then aggregated at the gateway server at the farmer's house. This aggregation could be using a Raspberry Pi or a more compute-heavy mobile GPU-enabled, Jetson-class device. The \textit{second model} consists of an embedded testbed consisting of leading-edge embedded devices in the Jetson class of mobile GPUs. 
Specifically, three embedded devices are currently used in our embedded testbed to benchmark heavy video object detection algorithms, useful for applications in our living lab. The devices are the \textit{Jetson TX2}, \textit{Jetson Xavier NX}, and \textit{Jetson AGX Xavier}, with different levels of CPU, GPU, and memory capacity. The relation between their computational capacities is Jetson TX2 $<$ Jetson Xavier NX $<$ Jetson AGX Xavier. \textit{Finally}, we enable opportunistic data ferrying, using drones. This is for traffic that is delay tolerant. The drones perform fly-by over the farm and are able to detect sensor feeds beamed to them from the ground sensor nodes (or sensor nodes on poles) in the living testbed. This enables them to carry high-bandwidth video, sound, or image feeds from the static nodes to the gateway node(s). The drones can also be used for temporal surveillance of the fields where gathered image feeds (over time intervals) can be used for a temporal health scan of the fields. In our current implementation, we use LoRa and LoRaWAN gateways for aggregating the multi-modal sensor data. In alternate settings, these could be replaced by NB-IoT with some infrastructure and recurring costs.

Our living lab testbed across different scenarios is an enabler for usable data collection and analytics from farms, even small-scale farms. This will require a careful adaptation of wireless, database, and ML technologies for the digital agriculture scenario and orchestration of these individual technology elements into a usable and low-cost end-to-end system, as described in Lattice~\cite{chaterji2021lattice}. There is momentum behind this through projects like FarmBeats, AirBand technologies, and Open Ag Data Alliance. 
We hope that our work by providing an exemplar instantiation will accelerate this movement.



\bibliographystyle{unsrt} 
\bibliography{references}

\end{document}